\documentstyle[epsfig]{mn}

\title{Dynamical Evolution of Rotating Stellar Systems: III. The Effect of
Mass Spectrum}

\author{ Eunhyeuk Kim$^{1,2}$\thanks{e-mail: ekim@cfa.harvard.edu}, 
Hyung Mok Lee$^1$\thanks{e-mail: hmlee@astro.snu.ac.kr}, and Rainer Spurzem$^3$
\thanks{e-mail: spurzem@ari.uni-heidelberg.de}\\
$^1$Astronomy Program, SEES, Seoul National University, Seoul 151-742, Korea\\
$^2$Harvard-Smithsonian Center for Astrophysics, 60 Garden Street, Cambridge,
MA 02138, USA\\
$^3$Astronomisches Rechen-Institut, M\"onchhof-Strasse 12-14, 
69120 Heidelberg, Germany}

\begin{document}
\maketitle

\begin{abstract}
We have studied the dynamical evolution of rotating star clusters with
mass spectrum using a Fokker-Planck code. As a simplest multi-mass
model, we first investigated the two-component clusters. 
Rotation is found to accelerate the dynamical evolution
through the transfer of angular momentum outward, as well as from the high
masses to the low masses. 
However, the
degree of acceleration depends sensitively on the assumed initial
mass function since dynamical friction, which generates mass segregation,
also tends to accelerate the evolution, and the combined effect of both
is not linear or multiplicative.
As long as dynamical friction dominates in the competition with
angular momentum exchange the heavy masses lose random energy and angular
momentum, sink towards the centre, but their remaining angular momentum
is sufficient to speed them up rotationally. This is gravo-gyro instability.
As a consequence, 
we find that the high mass stars in the central
parts rotate faster than low mass stars. This leads to the
suppression of mass segregation compared to the non-rotating
clusters. From the study of multi-component models, we observe
similar trends to the two-component models in almost all
aspects. 
The mass function changes less drastically for clusters
with rotation. Unlike non-rotating clusters, the mass function
depends on $R$ and $z$.
Our models are the only ones that can predict mass function and 
other quantities to be compared with new observations.
\end{abstract}
\begin{keywords}
celestial mechanics, stellar dynamics -- globular clusters:
general
\end{keywords}

\section{Introduction}
\medskip

Since the pioneering studies of Lupton \& Gunn (1987) and its subsequent
applications to fit actual globular cluster observations it is evident
that galactic globular clusters exhibit some degree of rotation, and that
there is also a consistent amount of observed flattening of their shapes
(White \& Shawl 1987). More recently, van Leeuwen et al. (2000) and
Anderson \& King (2003) measure globular cluster rotation in proper motion,
in order to derive the true direction of its rotational axis.
Though not dominant the amount of rotational 
energy could also not be neglected on the other hand side, as was
seen by the models of Einsel \& Spurzem, which showed
that even a moderate fraction of rotational energy in a cluster leads
to significantly faster core evolution. 

This is the third in a series of studies on the dynamical evolution 
of rotating stellar system by
using orbit-averaged 2D Fokker-Planck (FP) models that include
the effects of initial rotation. In the 
previous two papers, pre-collapse evolution 
(Einsel \& Spurzem 1999, Paper I)
and the evolution after core-collapse (Kim et al. 2002, Paper II) of 
rotating clusters composed of
equal mass stars were studied. In the present study we explore the 
dynamical evolution of the rotating stellar
systems with mass spectrum as a natural extension to the previous
works. We expect that the exchange of angular momentum between different
mass species would significantly affect the course of dynamical
evolution, as the energy exchange is known to play important role in the 
evolution of multi-mass clusters. 

%

Direct integration of the Fokker-Planck equation is used as 
a statistical method. Comparisons
between results obtained with FP method and results from $N$-body 
simulation show that the approximations and
assumptions which were used in FP models are reasonable, but need
to be checked carefully by comparison of different methods.
The simplest implementation of an FP model is a one-dimensional (1D) FP model 
where the distribution function $f$ is assumed to depend only on the energy 
$E$ of stars. All physical properties depend on the distance from the 
cluster center only (Cohn 1980; Lee, Fahlman \& Richer 1991).
The use of 1D models is inspired by 
the fact that the shape of globular clusters is approximately spherical, but 
it ignores the anisotropy of the velocity dispersion. 
Observations of globular clusters and theoretical models suggested
that there exists a difference in the velocity dispersion
between radial and tangential direction, especially for stars in the
outer regions of the clusters (Takahashi 1995, 1996, 1997). 
The two-dimensional FP model where distribution function depends on energy and
angular momentum with spherical geometry has been pioneered by Cohn (1979) but
studied extensively only recently because of difficulties in numerical 
integration of the equations (Takahashi, Lee \& Inagaki 1997, Takahashi \& Lee
2001). 
In our case, modelling axisymmetric systems with only two integrals
requires a careful check in particular. Any third integral is
neglected here completely, which means that the diffusion properties
of orbits in the axisymmetric potential are treated as a function of
two integrals only.

Although there are successes in previous FP models to explain the 
dynamical evolution of star clusters, only few of
them considered the natural and important physical property of
the existence of an initial angular momentum in the star cluster.
Kim et al. (2002) have reviewed briefly the previous studies where the 
initial rotation is considered. They presented 
the first post-core collapse studies on the evolution of rotating stellar 
systems and found that the global shape of the
rotational structure of the globular cluster changes little, though 
the strength of the rotation (measured using the magnitude of
the rotational velocity or the $z$-component of the angular momentum) 
decreases continuously with time due to the outward transfer of the angular 
momentum. An incorporation of mass loss and enhanced two-body relaxation 
processes accelerated the evolution of the star cluster not only 
in core-collapse time, but also the dissolution time of the cluster.
They found an approach to self-similar evolution in late
core-collapse.

The previous studies where the initial rotation is considered, 
assumed a star cluster with equal mass stars.
Inclusion of the initial mass spectrum known to
change the evolution of star clusters significantly, shortening of
core-collapse times for example (Lee, Fahlman \& Richer 1991). 
We also expect that there may be important physical processes between different 
mass species concerning the exchanges of the angular momentum. Until now, 
no study has been done on  rotating star clusters
with mass spectrum except for one preliminary work by 
Spurzem \& Einsel (1998). In this paper, thus we present the study of
dynamical evolution of the rotating star cluster with the initial mass
function (IMF hereafter). 
We first investigate the evolution of two-component models as a simplest
extension of the single mass models. Then we extend our study to the
multi-mass models represented by ten different mass species. 
The difference of our models as compared to the preliminary work
by Spurzem \& Einsel (1998) is that first our improved code includes
more accurate numerical integration and discretisation procedures as
described in Paper II, and here we do an extensive parameter study,
which did not exist before.

The outline of the present paper is as follows;
In section 2, the FP equations for the multi-mass system and 
the initial models are presented. We concentrate on the evolution up to 
the core collapse of two component models in section 3 and
multi-mass models in section 4. We further discuss the evolution
beyond the core collapse for both two-component and multi-component models
section 5. We summarize our main results in section 6.

\section{The Models}

\subsection{Fokker-Planck Equations}

We have constructed a computational scheme to study the dynamical evolution 
of rotating stellar systems with mass spectrums in detail with higher accuracy
both for pre- and post-collapse. The framework of the method is essentially an
extension of the method used in Paper II. The multi-mass 
FP equation under
a fixed potential can be written in a flux-conserving form as follows:

\begin{equation}
P(E,J_z){\partial f_i\over \partial t} = - {\partial F_{E_i}\over \partial E} - {\partial F_{J_{z,i}}\over \partial J_z}
\end{equation}
where $P(E,J_z)$ is the phase space volume accessible for
stars with $E$ and $J_z$, and 
$f_i$, $F_{E_i}$ and $F_{J_{z,i}}$ are the distribution
function and the particle flux in energy ($E$) and $z$-component of 
angular momentum ($J_z$) of the i-th component, respectively. 
The expression for phase space volume is given in Paper I. 
The particle
fluxes $F_{E_i}$ and $F_{J_{z,i}}$ can be expressed as follows:

\begin{eqnarray}
-F_{E_i} = D_{EE,i} {\partial f_i\over \partial E} + D_{EJ_z,i} 
{\partial f_i\over \partial J_z} + D_{E,i} f_i, \nonumber \\
\\
-F_{J_{z,i}} = D_{J_zJ_z,i} {\partial f_i\over \partial J_z} + D_{J_zE,i} 
{\partial f_i\over \partial E} + D_{J_{z,i}} f_i. \nonumber
\end{eqnarray}
where $D_{EE,i}$, $D_{EJ_z,i}$, etc are the diffusion coefficients and 
are given in the Appendix of Paper I for single component models. 
The extention to the multi-mass models can be simply done by applying
the distribution function of the $i$-th component in the expression of
these coefficients.

It is necessary to add energy source to explore the evolution after 
core-collapse. Primordial binaries and massive stars are important 
energy sources, since they can delay significantly the core-collapse 
time and change the early evolution of stellar
systems (Gao et al. 1991, Giersz \& Spurzem 2000). However,
the specification of primordial binaries could introduce
many more model parameters. The treatment of such binaries in
FP code is not so trivial either. Since we are mainly concerned with the
general effects of rotation on the dynamical evolution of star
clusters, we limit ourselves to simplest cases.
As a way to follow the dynamical evolution beyond the core collapse,
we have only considered the
heating effect of binaries formed by three-body processes. 
Heating formula for three-body binaries 
of Lee, Fahlman \& Richer 
(1991) is used in our models (see also Takahashi 1997 for
the discussion of this formula).
In their formulation, the total heating rate per unit volume is given by

\begin{equation}
\dot{E}_{tot} = C_bG^5 \left( \sum_i {{n_i m_i^2} \over \sigma_i^3} 
\right)^3\sigma_0^2
\end{equation}
where $n_i$, $m_i$, and $\sigma_i$ are the number density, stellar mass, and
one-dimensional velocity dispersion, respectively, of the $i$-th component, and
$\sigma_0$ is the mass density-weighted central one-dimensional 
velocity dispersion.
The standard value of $C_b=90$ is used for present study. Heating rate for each
mass component is obtained by distributing the total heating rate such that

\begin{equation}
\dot{E}_{tot,i} = {\rho_i\over \rho_{tot}} \dot{E}_{tot}
\end{equation}
where $\rho_i$ and $\rho_{tot}$ represent mass density of each component 
and total density. Giersz \& Spurzem (1997) have demonstrated that such a
simple approximations work very well for modelling the post-collapse
evolution of globular clusters.

\subsection{Multi-Mass Initial Models}
We employ the rotating King models as initial models following Lupton \&
Gunn (1987). We assume that there is no initial mass segregation among
different mass species.  Our present initial models are,
therefore simple extension of the initial models used in Papers I and II. 
These models are characterized by two parameters: dimensionless central 
potential $W_0$ and rotational parameter $\omega_0$. We have studied 
the evolution of clusters with $W_0=6$ and $W_0=3$, respectively. 
Model clusters are assumed to rotate in circular orbits around
mother galaxy, so that the mean density within the tidal radius ($r_t$) 
remains a constant throughout the evolution. We examined the evolution 
of clusters with various amount of the
initial rotations (see Tables 1 and 2 
for the list of the initial models
of two component, and multi-component, respectively).

First, we considered the  models with two-component mass species. 
Two-component models are expected to provide us with understanding the 
essential features of the evolution of the multi-component clusters 
(for example, the transfer of angular momentum between
two mass components). We have examined the wide range of individual stellar
mass ratio $\mu := m_2/m_1$ and total mass ratio $M_1/M_2 = (1-q)/q$,
where $q:=M_2/(M_1+M_2)$. The quantities $\mu$ and $q$ can be compared
with the work of Khalisi (2002) and Fregeau et al. (2002).

Next, we considered clusters which have continuous mass spectra. 
We choose simple power-law mass function for convenience. The number of 
stars in a mass interval ($m, m+dm$) is given by

\begin{equation}
dN(m) = C m^\alpha dm, \ \ \ m_{\rm min} \le m \le m_{\rm max}
\end{equation}
where $m_{\rm min}$ and $m_{\rm max}$ denote the minimum and the maximum 
masses, respectively. The dynamic range of mass spectrum, expressed as 
$m_{\rm max}/m_{\rm min}$ is 10 for present study.

We have studied the evolution of clusters with three different
shapes of mass function: $\alpha$=  $-1.20, -2.35$ and $-3.50$, where 
$\alpha=-2.35$ represents Salpeter-type mass function. We list the 
parameters of the models
with continuous mass spectrum in Tables 1 and 
2. For models of post core-collapse, we consider
only a model with a mass function with $\alpha=-2.35$.

The number of mass groups used in present
study is ten. The mass of each group is assigned such that

\begin{equation}
m_i = m_{\rm min} \left({ m_{\rm max}\over m_{\rm min} }\right)^{(i-{1\over 2})/10},  
(i=1,2,...10).
\end{equation}
The total mass of $i$th mass group is

\begin{equation}
M_i = \int\limits_{m_{i-{1\over 2}}}^{m_{i+{1\over 2}}} N(m) \, m \, dm.
\end{equation}

\begin{table}
\label{tab3-1}
\centering
\begin{minipage}{120mm}
\caption{Initial Models of clusters with $W_0=6$.}
\vspace{1truemm}
\begin{tabular}{@{}ccccccc@{}} \hline\hline
Model & $m_2/m_1$ & $M_1/M_2$ & $\alpha_0$ & $\omega_0$ & $T_r/T_k(0) $ & Phase \\
& & & & & [\%]& \\ [3pt] \hline\hline
    &      &           &     &   0.0     &   0.00  &  Pre  \\
    &      &           &     &   0.2     &   3.40  &  Pre  \\
    &      &           &     &   0.3     &   7.06  &  Pre  \\
    &      &           &     &   0.4     &  11.35  &  Pre  \\
    &      &           &     &   0.6     &  20.17  &  Pre  \\
M2A & 2    &      5    & $-$ &   0.8     &  27.96  &  Pre  \\
    &      &           &     &   1.0     &  34.23  &  Pre  \\
    &      &           &     &   1.2     &  39.13  &  Pre  \\
    &      &           &     &   1.4     &  42.93  &  Pre  \\
    &      &           &     &   1.6     &  45.88  &  Pre  \\[1pt] \hline
    &      &           &     &   0.0     &   0.00  &  Pre  \\
M2B & 2    &     10    & $-$ &   0.3     &   7.06  &  Pre  \\
    &      &           &     &   0.6     &  20.17  &  Pre  \\[1pt] \hline
    &      &           &     &   0.0     &   0.00  &  Post \\
M2C & 5    &     10    & $-$ &   0.3     &   7.06  &  Post \\
    &      &           &     &   0.6     &  20.17  &  Post \\[1pt] \hline
    &      &           &     &   0.0     &   0.00  &  Pre  \\
M2D & 5    &    100    & $-$ &   0.3     &   7.06  &  Pre  \\
    &      &           &     &   0.6     &  20.17  &  Pre  \\[1pt] \hline
    &      &           &     &   0.0     &   0.00  &  Pre  \\
M2E & 10   &     10    & $-$ &   0.3     &   7.06  &  Pre  \\
    &      &           &     &   0.6     &  20.17  &  Pre  \\[1pt] \hline
    &      &           &     &   0.0     &   0.00  &  Pre  \\
M2F & 10   &    100    & $-$ &   0.3     &   7.06  &  Pre  \\
    &      &           &     &   0.6     &  20.17  &  Pre  \\[1pt] \hline
    &      &           &         & 0.0   &   0.00  &  Pre  \\
MCA & $-$  &   $-$     & $-1.20$ & 0.3   &   7.06  &  Pre  \\
    &      &           &         & 0.6   &  20.17  &  Pre  \\[1pt] \hline
    &      &           &         & 0.0   &   0.00  &  Post \\
MCB & $-$  &   $-$     & $-2.35$ & 0.3   &   7.06  &  Post \\
    &      &           &         & 0.6   &  20.17  &  Post \\[1pt] \hline
    &      &           &         & 0.0   &   0.00  &  Pre  \\
MCC & $-$  &   $-$     & $-3.50$ & 0.3   &   7.06  &  Pre  \\
    &      &           &         & 0.6   &  20.17  &  Pre  \\[1pt] \hline
\end{tabular}

\vspace{2truemm}
$T_r$: total rotational energy

$T_k$: total kinetic energy 
\end{minipage}
\end{table}


\begin{table}
\label{tab3-2}
\centering
\begin{minipage}{120mm}
\caption{Initial Models of clusters with $W_0=3$.}
\vspace{1truemm}
\begin{tabular}{@{}ccccccc@{}} \hline\hline
Model & $m_2/m_1$ & $M_1/M_2$ & $\alpha_0$ & $\omega_0$ & 
$T_r/T_k(0) $ & Phase\\
& & & & & [\%] & \\[3pt] \hline\hline
    &      &           &     &   0.0     &   0.00  &  Pre  \\
    &      &           &     &   0.4     &   1.89  &  Pre  \\
M2A & 2    &      5    & $-$ &   0.8     &   6.96  &  Pre  \\
    &      &           &     &   1.2     &  13.84  &  Pre  \\
    &      &           &     &   1.5     &  19.37  &  Pre  \\
    &      &           &     &   1.6     &  21.20  &  Pre  \\[1pt] \hline
    &      &           &     &   0.0     &   0.00  &  Pre  \\
M2B & 2    &     10    & $-$ &   0.8     &   6.96  &  Pre  \\
    &      &           &     &   1.5     &  19.37  &  Pre  \\[1pt] \hline
    &      &           &     &   0.0     &   0.00  &  Post \\
M2C & 5    &     10    & $-$ &   0.8     &   6.96  &  Post \\
    &      &           &     &   1.5     &  19.37  &  Post \\[1pt] \hline
    &      &           &     &   0.0     &   0.00  &  Pre  \\
M2D & 5    &    100    & $-$ &   0.8     &   6.96  &  Pre  \\
    &      &           &     &   1.5     &  18.37  &  Pre  \\[1pt] \hline
    &      &           &     &   0.0     &   0.00  &  Pre  \\
M2E & 10   &     10    & $-$ &   0.8     &   6.96  &  Pre  \\
    &      &           &     &   1.5     &  19.37  &  Pre  \\[1pt] \hline
    &      &           &     &   0.0     &   0.00  &  Pre  \\
M2F & 10   &    100    & $-$ &   0.8     &   6.96  &  Pre  \\
    &      &           &     &   1.5     &  19.37  &  Pre  \\[1pt] \hline
    &      &           &         & 0.0   &   0.00  &  Pre  \\
MCA & $-$  &   $-$     & $-1.20$ & 0.8   &   6.96  &  Pre  \\
    &      &           &         & 1.5   &  19.37  &  Pre  \\[1pt] \hline
    &      &           &         & 0.0   &   0.00  &  Post \\
MCB & $-$  &   $-$     & $-2.35$ & 0.8   &   6.96  &  Post \\
    &      &           &         & 1.5   &  19.37  &  Post \\[1pt] \hline
    &      &           &         & 0.0   &   0.00  &  Pre  \\
MCC & $-$  &   $-$     & $-3.50$ & 0.8   &   6.96  &  Pre  \\
    &      &           &         & 1.5   &  19.37  &  Pre  \\[1pt] \hline
\end{tabular}

\end{minipage}
\end{table}

If the initial density distribution and the initial mass function including
$m_{min}$ and $m_{max}$ are fixed, 
the only free parameter for the model is the number of stars
(or total mass if the masses of stars are given in physical units).
The general behavior of the core collapse does not depend on the
number of stars. However, the core stops to collapse and begins to 
expand when the central density exceeds a certain value which depends on
the total number of stars in the cluster.
If $N$ is too small, we cannot apply the Fokker-Planck method to follow 
the evolution of the star cluster.
The development of gravothermal oscillation occurs 
for $N$ much larger than 10,000 because the post-collapse cluster
becomes gravothermally unstable (Goodman 1987). The use of larger 
$N$ also takes more computational time because the post-collapse expansion
begins at much higher central density. As a compromise,
we have used $N = 10,000$ for models in the present study.  

The number of stars used here is quite small compared to that 
of globular clusters. According to the recent comparisons between 
direct $N$-body models and the Fokker-Planck
models 
(Takahashi \& Portegies Zwart 1998,
Portegies Zwart \& Takahashi 1999) the evolution of the total mass until total dissolution depends on
the particle number since the mass loss process involves the crossing time rather than the relaxation time.
Nevertheless we restrict ourselves here to post-collapse models 
using relatively small $N$, because this
speeds up the computational time, and we are mainly interested 
in the interplay between rotation,
relaxation and tidal mass loss. $N = 10,000$ is large enough
to study the effect of the core bounce due to the formation of binaries.
Aside from the actual values of the central density, the general
behavior of the post-collapse evolution does not sensitively depend
on initial $N$. Note that there will be rapid core oscillation, which is
beyond the scope of the present study, if we use realistic particle numbers.

\subsection{Tidal Boundary}
All models studied in the present work are tidally limited.
Realistic treatment of tidal field is possible only for direct N-body
calculations. The simplest assumption for the tidal boundary is to
assume a spherical shape whose radius is determined by the requirement
of constant mean density within this radius. The actual shape is not
exactly spherical and the tidal field varies with time as the cluster
moves along non-circular orbit in non-spherical Galactic potential. 
Nevertheless, we assumed a constant mean density within the tidal
radius for simplicity.

We have applied the ``energy condition'' for the removal of stars
(see Takahashi, Lee \& Inagaki 1997 for details of tidal conditions).
Takahashi \& Portegies Zwart (1998) found that the ``apocenter condition''
reproduces the result of the N-body simulation better for small $N$
models, but the difference between different conditions diminishes as
$N$ becomes larger. Since our distribution functions do not depend on
total angular momentum, we could not apply the apocenter condition.
However, we emphasize that our choice of $N=10,000$ is simply 
to reduce the
computational time. Although the parameters for the central parts 
(central density, core radius and central velocity dispersion) are
sensitive to the $N$, the global evolution (such as $M(t)$, $r_h(t)$) do
not depend on $N$ as long as we use large $N$ approximation
for the tidal conditions. We could consider our results to be appropriate
for systems with very large $N$, except for the core properties.

\section{Pre-collapse Evolution of Two-component Models}

Under the context of our models of two-body relaxation, rotation, and
binary heating the clusters always go through the core-collapse and begins 
the post-collapse expansion.
Some aspects of the evolution differ between pre- and post-collapse phases.
Thus we first discuss the pre-collapse evolution of two-component 
models first. In the study of the pre-collapse phase only,
we have ignored the heating by binaries.


For clusters with multiple mass, the energy exchange between different
mass species causes mass segregation which takes place in a time
scale shorter than the single-mass core-collapse. Once the core is 
dominated by the massive component whose velocity dispersion is lower than
average, the core collapse proceeds more rapidly than the
single component clusters. 
The rotation is also known to accelerate the core-collapse process since it
provides additional mechanism of driving the catastrophic collapse.
In two-component clusters, these two processes compete with each other. 
We now investigate how the rotation affects the course of
core-collapse of two-component models.

\subsection{Central density and core-collapse time}
\begin{figure}
\epsfig{figure=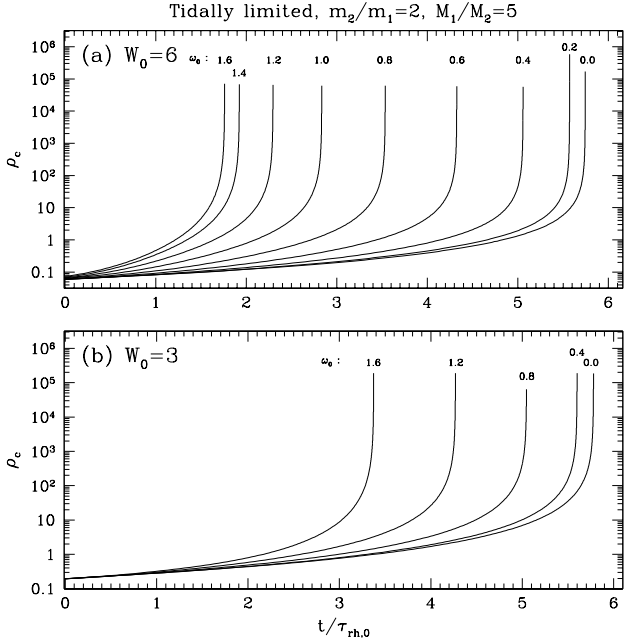, height=0.50\textwidth, width=0.50\textwidth}
\vspace{-3mm}
\caption{Time evolution of the central density for the cluster model with mass function M2A}
\label{fig3-1}
\end{figure}

We have shown the evolution of 
central density for the model M2A with different initial central potential,
$W_0=3$ and $6$ in Fig. \ref{fig3-1}. 
he time is expressed in units of the 
initial half-mass relaxation 
time($\tau_{rh,0}$), and the central density in units of $M_0/r_{c,0}^3$, 
where $M_0$ and $r_{c,0}$ are the initial mass of the cluster and the initial
King's core radius, respectively. An expression for the initial half-mass relaxation time is taken
from Spitzer \& Hart (1971),

\begin{equation}
\tau_{rh,0} = {{0.138 N_0^{1/2} r_{h,0}^{1/2}} \over { G^{1/2} {\bar{m}}^{1/2} \rm {ln}\, \Lambda }}
\end{equation}
where $N_0, r_{h,0}, G, {\bar{m}}$ and $\rm{ln}\, \Lambda$ 
denote the initial total number of stars the
initial half-mass radius, gravitational constant, the mean mass of 
the particles and the Coulomb logarithm, respectively. Since the rotating
clusters are not spherically symmetric, some cautions should be
taken in defining the half-mass radius. It is defined as the
effective radius of concentric spheroid within which the enclosed
mass becomes half of the total mass (see eq. [22] of Einsel \&
Spurzem 1999).

\begin{figure}
\epsfig{figure=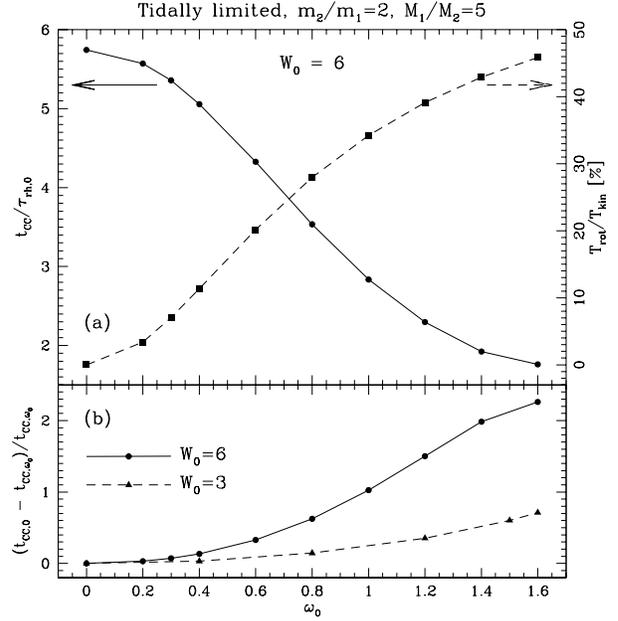, height=0.50\textwidth, width=0.50\textwidth}
\vspace{-3mm}
\caption{(a) Run of core collpase time measured in initial half mass relexation time as a function 
of the initial rotation rotational parameter ($\omega_0$) for a series of models
with a mass function M2A and $W_0=6$ (solid line). A ratio between the initial rotational energy
and the total kinetic energy for the same models (dashed line). (b) The degree of the acceleration
of core-collapse defined by eq. \ref{deg_acc} for 
model with the central potential $W_0 = 6$ (solid line) and $3$ (dashed line), respectively.}
\label{fig3-2}
\end{figure}

It is shown clearly that the clusters with higher initial rotation reach 
core-collapse earlier than the cluster without initial rotation. In order to
quantify the amount of acceleration of the evolution, we define the
``degree of acceleration''  as follows:
\begin{equation}
d_{acc} \equiv {t_{cc}(\omega=0)-t_{cc}(\omega )\over t_{cc}(\omega)},
\label{deg_acc}
\end{equation}
where $t_{cc}$ is the time to reach the core-collapse. Obviously, $d_{acc}$
depends on $\omega$ for a given $W_0$ and a mass function. We have 
listed $d_{acc}$ in the last column of Tables 
3 and 4, and have shown them
in Fig. \ref{fig3-2}(b).

Fig. \ref{fig3-2}(a) shows the run of the core-collapse times as a 
function of the initial rotation (solid line) and the ratio between 
initial total rotational energy due to the rotation and initial 
total kinetic energy (dashed line) for model with mass 
function of M2A and the initial central potential $W_0=6$. The
ratio between rotational energy and kinetic energy of the model 
with initial rotation $\omega_0 = 1.6$
is $\sim 46\%$, just below the dynamically unstable criterion of 
$T_r/T_k = 0.5$, where $T_r$ and
$T_k$ denote the total rotational energy and total kinetic energy. 
The core-collapse times, which are measured in units of the initial half-mass 
relaxation time decrease as the initial degree
of rotation increases. The cluster with the initial rotation $\omega_0=0.6$ 
reaches the core collapse faster by a factor
of $\sim 1.3$ than the cluster without the initial rotation. The cluster 
with the initial rotation $\omega_0=1.6$, a 
model with the largest initial degree of rotation among current  
models, has a core-collase time of just $\sim 1/3$ of that
of the model without the initial rotation. 

\begin{figure}
\epsfig{figure=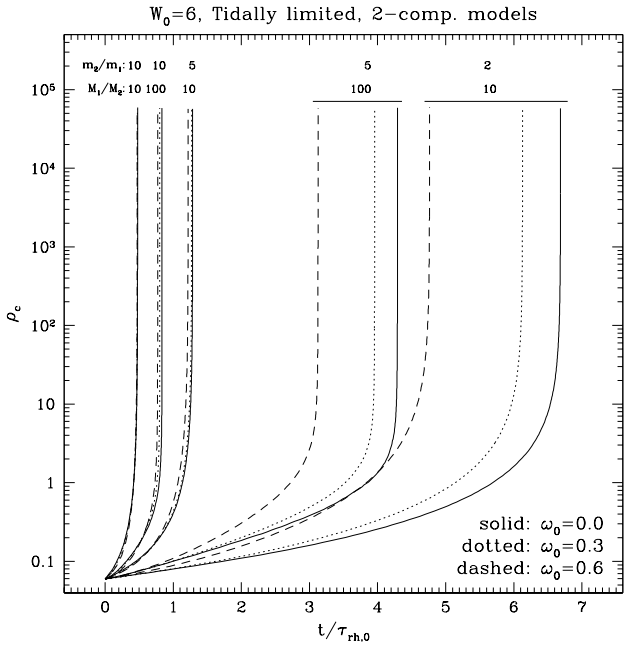, height=0.50\textwidth, width=0.50\textwidth}
\vspace{-3mm}
\caption{Time evolution of the central density for the cluster model with two-component mass functions and
$W_0=6$ except for a mass function of M2A.} 
\label{fig3-3}
\end{figure}

We have shown the time evolution of central density for models with 
different mass function except for mass function of M2A in Fig. \ref{fig3-3}. 
Solid lines represent the models without initial rotation. Dotted lines
and dashed lines represent the models with the initial degree of 
rotation $\omega_0=0.3$ and $0.6$, respectively.
We have chosen the values for the initial degree of rotation as 
the same values used in Paper II.
The mass ratios between high and low mass components($m_2/m_1$) and the 
ratios of total masses of these components ($M_1/M_2$) are written 
to distinguish the different mass functions.

In general the core collapse occurs earlier for more rapidly rotating models.
The degrees of acceleration also depends
sensitively on the assumed mass function.
For example, $d_{acc}$ of the model with the initial rotation $\omega_0=0.6$ 
and mass function of M2B is much larger than that of  
the model with mass function M2E, which has the same initial degree of 
rotation. The main difference between these two models is the ratio of
the individual mass: $m_2/m_1=2$ for M2B and 10 for M2E. The larger difference 
in individual mass means larger amount of energy exchange and thus the 
significant amount of acceleration of core collapse due to this effect
alone. There appears not enough room for the acceleration due to the initial
rotation for such cases. For a given $m_2/m_1$, the degree of
acceleration also depends on $M_1/M_2$. 
If the total mass in high mass component is relatively large (i.e., smaller
values for $M_1/M_2$), the acceleration due to rotation is small. Such trend
can be seen from the comparison between M2C and M2D or M2E and
M2F. 
For models with relatively small $M_1/M_2$, the acceleration due to
mass segregation is already significant and the role of rotation is 
relatively less important. 


\begin{figure}
\epsfig{figure=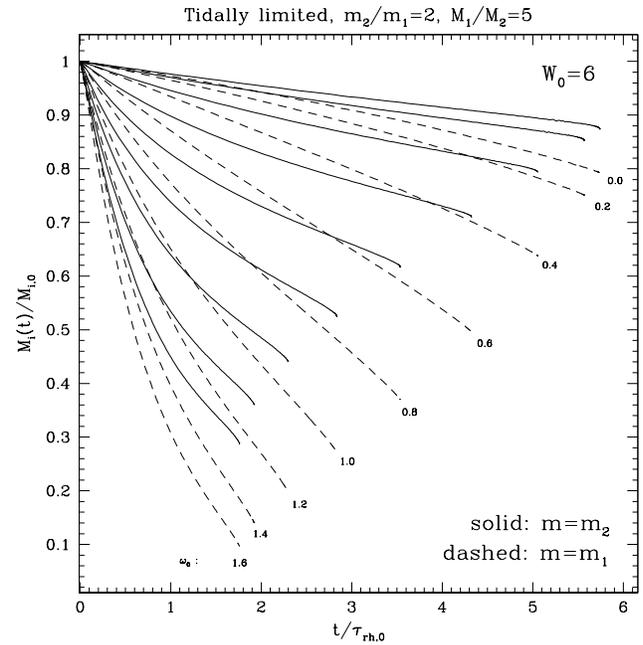, height=0.50\textwidth, width=0.50\textwidth}
\vspace{-3mm}
\caption{Evolution the total masses retained in a cluster for model with $W_0=6$ and mass
function M2A. The total mass of the individual component is normalized by its initial
total mass $M_{i,0}$. Solid lines represent the evolution for the massive stars and dashed lines 
for the less massive stars, respectively.}
\label{fig3-4}
\end{figure}

Time evolution of total mass of the clusters with mass function M2A is 
displayed in Fig. \ref{fig3-4}. The total mass of each mass component 
is normalized by their initial total mass $M_{i,0}$. The dashed lines
represent the total mass of lighter component and the solid lines 
for massive component, respectively. The total mass of lighter component 
decreases more rapidly than the that of massive stars, irrespective of 
the initial degree of rotations. The cluster with a initially higher 
rotation loses mass more efficiently than the clusters with lower
initial rotation. The higher moss-loss rate of 
cluster with the higher initial rotation was known in previous works 
for single mass models (Papers I and II). 
The ratio of total masses at a 
time of core-collapse between two mass components varies with different
initial degree of rotations. For the model with initial rotation 
of $\omega_0=1.6$, the total mass of the low mass component at the time of 
core-collapse is $\sim 10$ \% of its initial value. On the other hand, 
$\sim 30$ \% of the initial mass still remains for massive component. 
The ratio of total mass at a time of core-collapse between the high and low 
mass components for model without the initial rotation is $\sim 1.1$, 
which is much lower than the value obtained for cluster model
with the initial rotation $\omega_0=1.6$. It implies that the mass function 
of rotating cluster which has the same initial mass function will 
follow different evolution according to their initial rotation.
The evolution of mass function, especially the power law index 
$\alpha$, for models with initially power-law mass function is 
explained in detail in Sect. 5.4.

\subsection{Central velocity dispersion and central angular speed}

\begin{figure}
\epsfig{figure=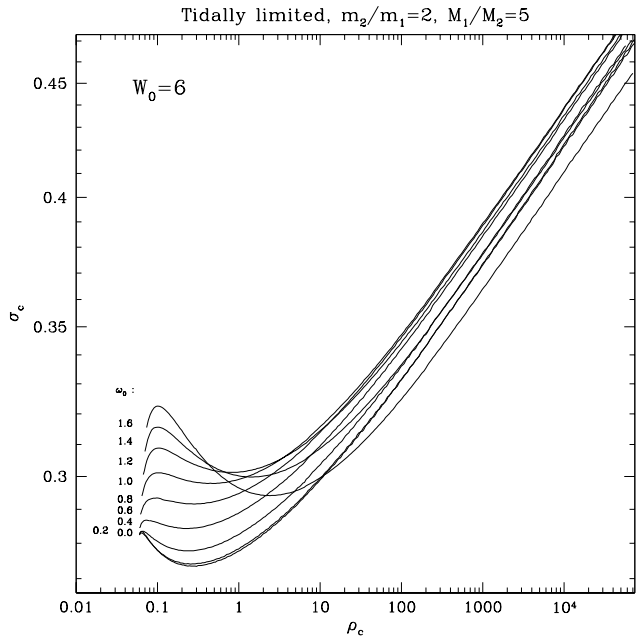, height=0.50\textwidth, width=0.50\textwidth}
\vspace{-3mm}
\caption{Evolution of the mass-weighted central 1D velocity dispersion with respect to the
central density for cluster models with $W_0=6$ and M2A for a mass function. The values of initial 
rotational parameter ($\omega_0$) are written ahead of the starting points of the evolution.}
\label{fig3-5}
\end{figure}

Einsel \& Spurzem (1999) argued in their single mass model that the 
acceleration of core-collapse time for models with the initial 
rotation compared to the model without the initial rotation is 
accompanied by a rapid increase of the central velocity dispersion in 
rotating models.  We have checked if
this is also true for two component models. Fig. \ref{fig3-5}
shows the evolution of mass density weighted central one-dimensional 
velocity dispersion ($\sigma_c$) for models with the central potential 
$W_0=6$ and mass function model of M2A. The initial rotational parameters 
($\omega_0$) are written in front of the starting points of evolution.

The evolution of $\sigma_c$ measured in the total central density ($\rho_c$) 
shows a rather complex behavior. For models with large initial rotation,
$\sigma_c$ increases in the very early phase of evolution and then decrease
until it reaches broad minimum. Eventually, $\sigma_c$ increases with
$\rho_c$ monotonically toward the core collapse. 

The initial rise of $\sigma_c$ is caused by the initial collapse and 
subsequent heating by contraction.
The decrease of $\sigma_c$ after the
local maximum is due to the loss of energy of high mass stars to low
mass stars. Since there are more low mass stars in the early phase in the
central parts, the mass-weighted velocity dispersion decreases.
The subsequent increase of $\sigma_c$ after the broad minimum occurs when
the central part is dominated by high mass stars. The stars in the
central parts lose energy to stars in the outer parts and the velocity
dispersion increases because of the negative specific heat.


The evolutionary path reaches self-similar 
collapse phase when the central density increases by $2 \sim 3$ orders
of magnitude. The evolution of self-similar collapse can be characterized 
by a parameter given by
 
\begin{equation}
\gamma = { { d\, \rm{ln}\, \sigma_c^2 } \over { d\, \rm{ln}\, \rho_c }}.
\end{equation}
For single mass system, 
Einsel \& Spurzem (1999) obtained an average value of $\gamma = 0.109$, 
independent of the initial degree of rotation. It is expected that 
$\gamma$ for present two component models is very similar to that 
of the single component system, 
since the central part of cluster is dominated by high mass stars.
We derive an average value of $\gamma = 0.105$, which is very similar to 
what was obtained for single mass model. 

Another feature of Fig. \ref{fig3-5} is the initial evolution 
of of $\sigma_c$ depends on the amount of initial rotation. 
The minima occur at higher central density for models with larger initial
rotation. The enhanced two-body relaxation process due to
rotation will eventually cause more rapid increase the central density.
This results in the acceleration of the core-collapse for
rapidly rotating systems. 

\begin{figure}
\epsfig{figure=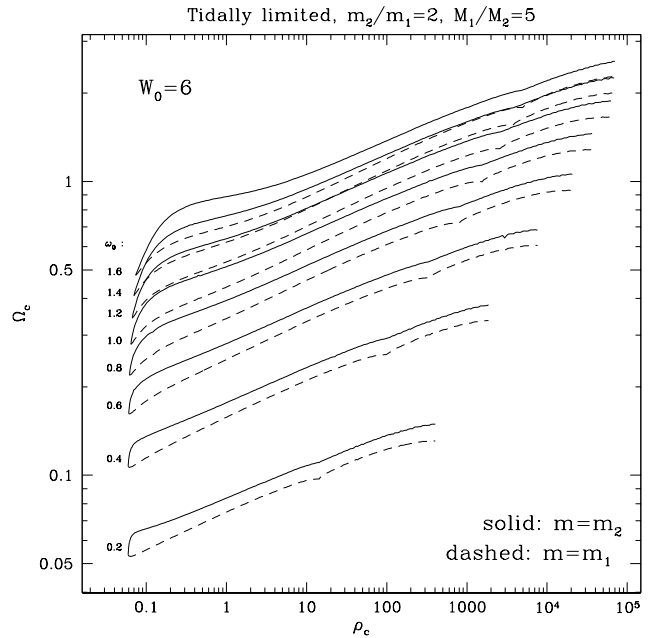, height=0.50\textwidth, width=0.50\textwidth}
\vspace{-3mm}
\caption{Evolution of the central angular speed of the individual mass species with respect to the
total central density for cluster models with $W_0=6$ and M2A for a mass function. The central
rotational speed of the massive stars (solid lines) increases very fast during the early evolutionary
stage, while that of the lighter stars (dashed lines) shows an moderate increase.}
\label{fig3-6}
\end{figure}

\begin{center}

\begin{table*}
\label{tab3-3}
\centering
\begin{minipage}{150mm}
\caption{Physical properties of cluster models with $W_0=6$.}
\vspace{1truemm}
\begin{tabular}{@{}cccccccc@{}} \hline\hline
Model & $\omega_0$ & $t_{CC}/\tau_{rh,0}$ & $M_{CC}/M_0$ & 
$t_{dis}/\tau_{rh,0}$ & $d_{acc}$ & $\gamma$ & $\delta$  \\[3pt] \hline\hline
    & 0.0     &   5.74   &  0.81   &  $-$ & 0.00 & 0.108 & $-$    \\
    & 0.2     &   5.57   &  0.78   &  $-$ & 0.03 & 0.114 & 0.112  \\
    & 0.3     &   5.36   &  0.72   &  $-$ & 0.07 & 0.106 & 0.114  \\
    & 0.4     &   5.06   &  0.66   &  $-$ & 0.14 & 0.106 & 0.113  \\
M2A & 0.6     &   4.33   &  0.53   &  $-$ & 0.33 & 0.105 & 0.111  \\
    & 0.8     &   3.54   &  0.41   &  $-$ & 0.62 & 0.105 & 0.108  \\
    & 1.0     &   2.83   &  0.31   &  $-$ & 1.03 & 0.105 & 0.108  \\
    & 1.2     &   2.30   &  0.24   &  $-$ & 1.50 & 0.105 & 0.111  \\
    & 1.4     &   1.92   &  0.18   &  $-$ & 1.99 & 0.105 & 0.109  \\
    & 1.6     &   1.76   &  0.13   &  $-$ & 2.26 & 0.106 & 0.107  \\[1pt] \hline
    & 0.0     &   6.69   &  0.78   &  $-$ & 0.00 & 0.106 & $-$    \\
M2B & 0.3     &   6.13   &  0.70   &  $-$ & 0.09 & 0.105 & 0.108  \\
    & 0.6     &   4.76   &  0.52   &  $-$ & 0.40 & 0.104 & 0.111  \\[1pt] \hline
    & 0.0     &   1.28   &  0.96   &  10.68 & 0.00 & 0.106 & $-$    \\
M2C & 0.3     &   1.27   &  0.93   &   7.74 & 0.01 & 0.105 & 0.090  \\
    & 0.6     &   1.22   &  0.82   &   4.36 & 0.05 & 0.104 & 0.092  \\[1pt] \hline
    & 0.0     &   4.29   &  0.88   &  $-$ & 0.00 & 0.101 & $-$    \\
M2D & 0.3     &   3.96   &  0.82   &  $-$ & 0.08 & 0.099 & 0.092  \\
    & 0.6     &   3.13   &  0.67   &  $-$ & 0.37 & 0.102 & 0.094  \\[1pt] \hline
    & 0.0     &   0.48   &  0.98   &  $-$ & 0.00 & 0.106 & $-$    \\
M2E & 0.3     &   0.47   &  0.97   &  $-$ & $<$0.01 & 0.105 & 0.090 \\
    & 0.6     &   0.47   &  0.90   &  $-$ & 0.01 & 0.104 & 0.084  \\[1pt] \hline
    & 0.0     &   0.84   &  0.98   &  $-$ & 0.00 & 0.099 & $-$    \\
M2F & 0.3     &   0.80   &  0.96   &  $-$ & 0.05 & 0.099 & 0.099  \\
    & 0.6     &   0.77   &  0.90   &  $-$ & 0.08 & 0.101 & 0.100  \\ \hline
    & 0.0     &   3.49   &  0.83   &  $-$ & 0.00 & 0.106 & $-$    \\
MCA & 0.3     &   3.39   &  0.74   &  $-$ & 0.03 & 0.104 & 0.095  \\
    & 0.6     &   2.83   &  0.54   &  $-$ & 0.23 & 0.103 & 0.099  \\ \hline
    & 0.0     &   1.98   &  0.92   &  9.60 & 0.00 & 0.104 & $-$    \\
MCB & 0.3     &   1.92   &  0.86   &  7.11 & 0.03 & 0.103 & 0.105  \\
    & 0.6     &   1.72   &  0.70   &  4.16 & 0.15 & 0.103 & 0.113  \\ \hline
    & 0.0     &   1.73   &  0.94   &  $-$ & 0.00 & 0.098 & $-$    \\
MCC & 0.3     &   1.68   &  0.90   &  $-$ & 0.03 & 0.096 & 0.105  \\
    & 0.6     &   1.55   &  0.77   &  $-$ & 0.12 & 0.096 & 0.103  \\ \hline
\end{tabular}

\vspace{2truemm}
$t_{CC}/\tau_{rh,0}$: core collapse time measured in the initial half-mass relaxation time \\
$M_{CC}/M_0$: mass of the cluster in unit of the initial total mass ($M_0$) at $t=t_{CC}$ \\
$t_{dis}/\tau_{rh,0}$: time in unit of $\tau_{rh,0}$ when the cluster dissolved completely \\
$\gamma, \delta$: see the text for the definitions.
\end{minipage}
\end{table*}
\end{center}


\begin{table*}
\label{tab3-4}
\centering
\begin{minipage}{150mm}
\caption{Initial Models of clusters with $W_0=3$.}
\vspace{1truemm}
\begin{tabular}{@{}cccccccc@{}} \hline\hline
Model & $\omega_0$ & $t_{CC}/\tau_{rh,0}$ & $M_{CC}/M_0$ & 
$t_{dis}/\tau_{rh,0}$ & $d_{acc}$ & $\gamma$ & $\delta$  \\[3pt] \hline\hline
    & 0.0  & 5.78 & 0.42 & $-$ & 0.00 & 0.107 & $-$   \\
    & 0.4  & 5.60 & 0.39 & $-$ & 0.03 & 0.106 & 0.110 \\
M2A & 0.8  & 5.05 & 0.33 & $-$ & 0.15 & 0.106 & 0.112 \\
    & 1.2  & 4.27 & 0.25 & $-$ & 0.35 & 0.106 & 0.113 \\
    & 1.5  & 3.61 & 0.18 & $-$ & 0.60 & 0.106 & 0.109 \\
    & 1.6  & 3.37 & 0.15 & $-$ & 0.71 & 0.106 & 0.111 \\[1pt] \hline
    & 0.0  & 6.21 & 0.41 & $-$ & 0.00 & 0.105 & $-$   \\
M2B & 0.8  & 5.37 & 0.33 & $-$ & 0.16 & 0.104 & 0.107 \\
    & 1.5  & 3.79 & 0.19 & $-$ & 0.64 & 0.104 & 0.113 \\[1pt] \hline
    & 0.0  & 1.98 & 0.76 & 4.17 & 0.00 & 0.105 & $-$   \\
M2C & 0.8  & 1.88 & 0.66 & 3.43 & 0.05 & 0.104 & 0.094 \\
    & 1.5  & 1.59 & 0.39 & 2.01 & 0.25 & 0.105 & 0.090 \\[1pt] \hline
    & 0.0  & 4.82 & 0.57 & $-$ & 0.00 & 0.104 & $-$   \\
M2D & 0.8  & 4.82 & 0.49 & $-$ & 0.18 & 0.103 & 0.095 \\
    & 1.5  & 2.80 & 0.34 & $-$ & 0.72 & 0.104 & 0.099 \\[1pt] \hline
    & 0.0  & 0.86 & 0.87 & $-$ & 0.00 & 0.105 & $-$   \\
M2E & 0.8  & 0.84 & 0.80 & $-$ & 0.02 & 0.104 & 0.084 \\
    & 1.5  & 0.76 & 0.56 & $-$ & 0.13 & 0.104 & 0.085 \\[1pt] \hline
    & 0.0  & 1.54 & 0.86 & $-$ & 0.00 & 0.102 & $-$   \\
M2F & 0.8  & 1.49 & 0.79 & $-$ & 0.03 & 0.101 & 0.087 \\
    & 1.5  & 1.37 & 0.60 & $-$ & 0.13 & 0.101 & 0.093 \\[1pt] \hline
    & 0.0  & 4.01 & 0.38 & $-$ & 0.00 & 0.104 & $-$   \\
MCA & 0.8  & 3.51 & 0.30 & $-$ & 0.14 & 0.105 & 0.106 \\
    & 1.5  & 2.51 & 0.15 & $-$ & 0.60 & 0.105 & 0.103 \\[1pt] \hline
    & 0.0  & 2.65 & 0.56 & 4.18 & 0.00 & 0.102 & $-$   \\
MCB & 0.8  & 2.42 & 0.45 & 3.48 & 0.09 & 0.101 & 0.142 \\
    & 1.5  & 1.83 & 0.23 & 2.31 & 0.45 & 0.103 & 0.139 \\[1pt] \hline
    & 0.0  & 2.48 & 0.70 & $-$ & 0.00 & 0.096 & $-$   \\
MCC & 0.8  & 2.29 & 0.58 & $-$ & 0.08 & 0.096 & 0.118 \\
    & 1.5  & 1.82 & 0.34 & $-$ & 0.36 & 0.096 & 0.135 \\[1pt] \hline
\end{tabular}

\end{minipage}
\end{table*}
Fig. \ref{fig3-6} displays the run of central angular 
speed ($\Omega_c$) as a function of the total central density for 
the model with the mass function of M2A. 
The solid lines and the dashed lines represent $\Omega_c$ of 
high mass and low mass components, respectively. It is 
shown clearly that $\Omega_c$ of the higher mass component 
increases very quickly only during early phase of evolution. For the models
rotating slowly ($\omega_0 = 0.2$ and $0.4$) $\Omega_c$ of lower 
mass component decreases by a small amount. For rapidly rotating clusters,
however, $\Omega_c$ increases with increase of
central density. 
The rapid increase of $\Omega_c$ of high mass component in the early
phase is a caused by rapid collapse of the high mass stars: as they
lose energy, they rotate faster.  

\begin{figure}
\epsfig{figure=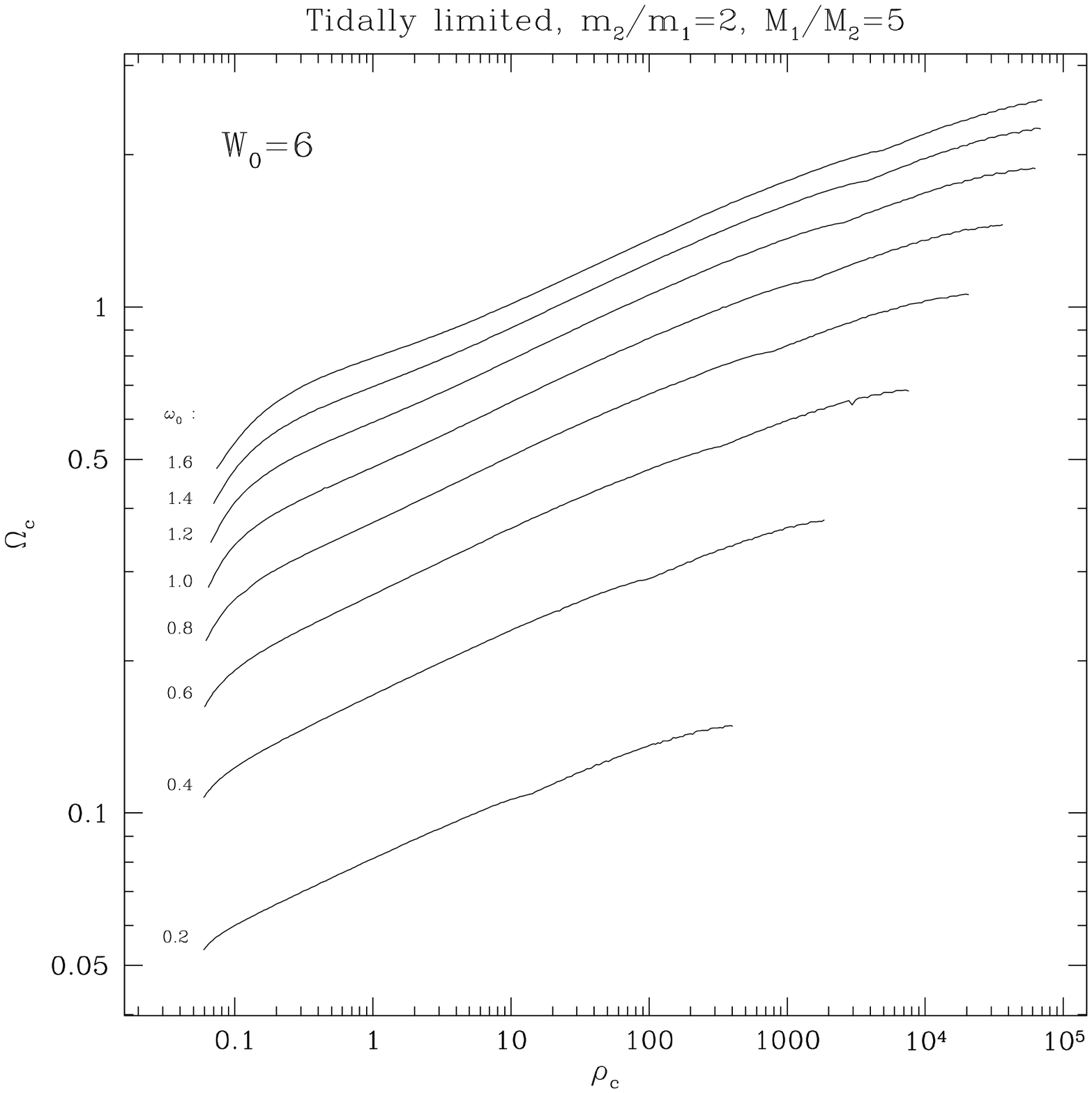, height=0.50\textwidth, width=0.50\textwidth}
\vspace{-3mm}
\caption{Evolution of the mass-weighted central angular speed with respect to the
central density for cluster models with $W_0=6$ and M2A for a mass function.}
\label{fig3-7}
\end{figure}

The mass weighted angular speeds in the core as a function of $\rho_c$ 
for the models in Fig. \ref{fig3-6} are shown in Fig. \ref{fig3-7}. During 
early evolution, $\Omega_c$ increases rapidly for models with large
rotation, then it grows slowly with increase of central density, 
reaching self-similar phase in evolutionary path.  Einsel \& Spurzem (1999) 
introduced the parameter $\delta$ which characterizes the self-similar
evolution of central angular speed as follows,

\begin{equation}
\delta = { { d\, \rm{ln}\, \Omega_c } \over { d\, \rm{ln}\, \rho_c } }.
\end{equation}

The calculated values of $\delta$ for the present two-component models are
listed in Table 3. On average, $\delta \approx 0.110$ with weak dependence 
on the initial rotation for model M2A. This value of $\delta$ 
is about twice of those of single mass models. 
Einsel \& Spurzem (1999) obtained $\delta = 0.06 \sim 0.08$ depending on 
the initial rotation for their model cluster with central potential $W_0=6$. 

\begin{figure}
\epsfig{figure=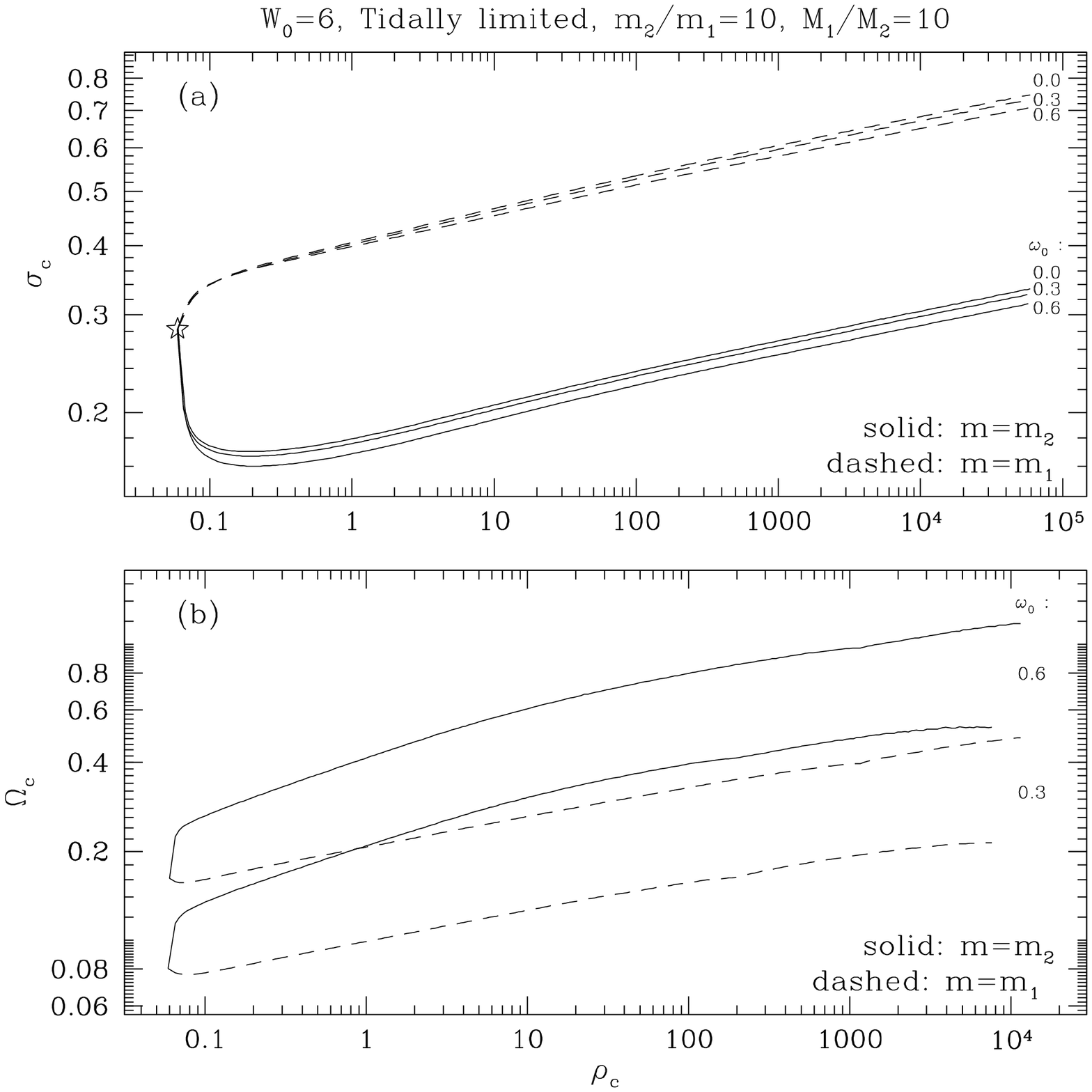, height=0.50\textwidth, width=0.50\textwidth}
\vspace{-3mm}
\caption{Evolution of the central velocity dispersion (a) and the central angular
speed (b) for the models with $W_0=6$ and a mass function M2E ($m_2/m_1=10, M_1/M_2=10$),
respectively. The start of the evolution is marked with starry mark in figure (a). We have
included models with the initial rotation of $\omega_0=0.0, 0.3$ and $0.6$. However, only the
rotating models are considered in $\Omega_c$ versus $\rho_c$ plot. The steep decreasing of
the central velocity dispersion and the steep rising for central angular speed of the massive
component are shown clearly. The loss of central velocity dispersion is compensated by the
quick increasing of angular speed.}
\label{fig3-8}
\end{figure}

The relationship between the central density, $\sigma_c$, and $\Omega_c$ 
for models with mass function M2E ($m_2/m_1=10, M_1/M_2=10$) are shown in 
Fig. \ref{fig3-8}. For $\sigma_c$ versus $\rho_c$, we have plotted the 
velocity dispersion of individual mass component for model without initial
rotation and models with $\omega_0 = 0.3$ and $0.6$, while only the rotating 
models are shown for
$\Omega_c$ versus $\rho_c$ plot. The starry mark in Fig. \ref{fig3-8}(a) 
represents the starting point of evolution for both mass components. 
The central velocity dispersion of high mass component decreases
dramatically, while that of low mass 
component increases slightly at early evolutionary stage due to the energy 
equipartition between two mass components. The decrease of 
the central velocity 
dispersion for model with higher initial rotation is larger than that of
the lower initial rotation. Inspection 
of  Fig. \ref{fig3-8}(a) reveals that the slope in log-log plot for massive 
component becomes $\gamma \approx 0.105$ regardless of degree of rotation.

\subsection{Rotational Velocity and Angular Momentum}

\begin{figure}
\epsfig{figure=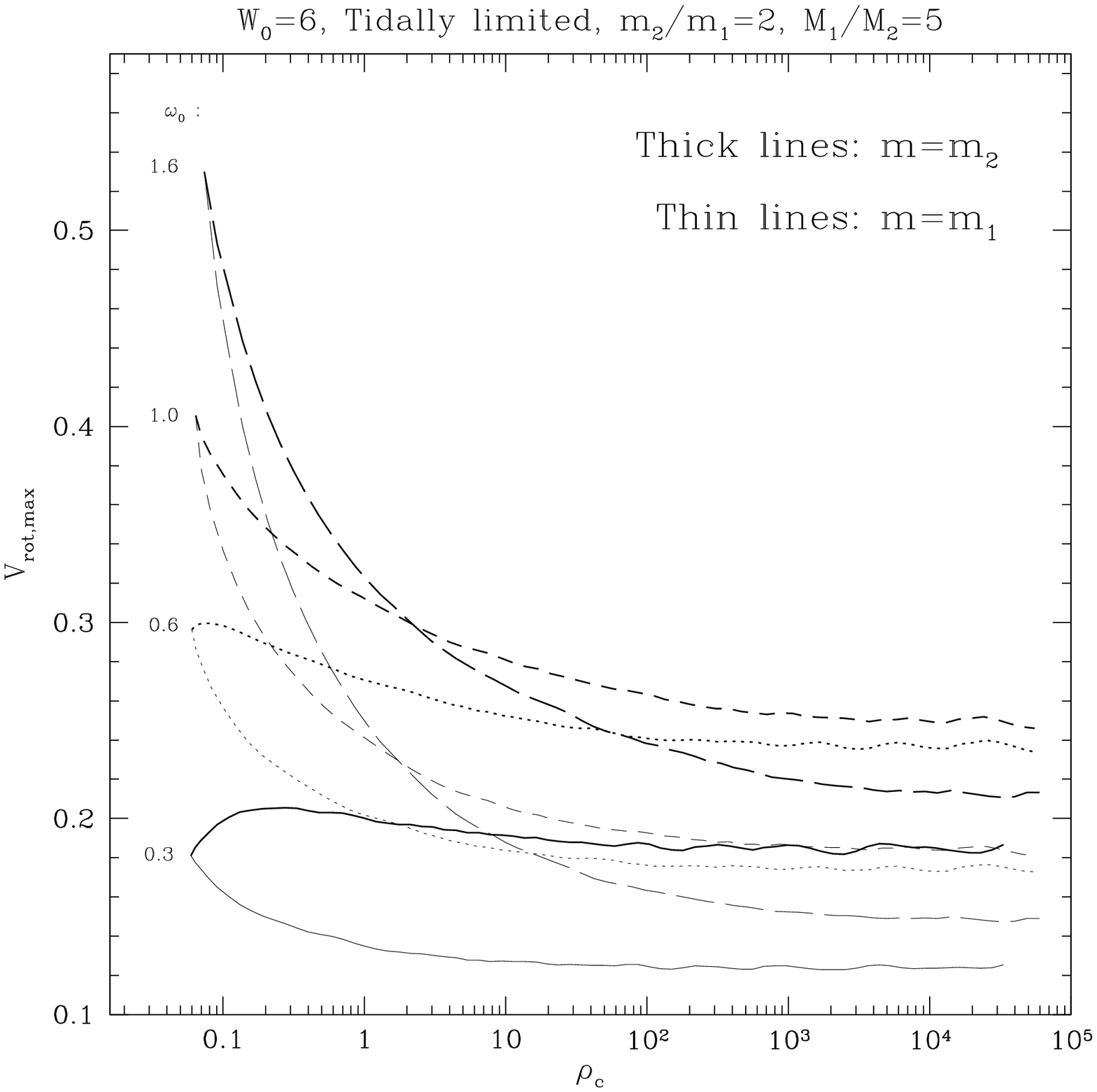, height=0.50\textwidth, width=0.50\textwidth}
\vspace{-3mm}
\caption{Evolution of the maximum rotational velocity at the equator according to the central
density for models with $W_0=6$ and mass function of M2A. We have shown only for four different
initial rotations out of 9 models. The thin lines and the thick lines represent the evolution of
the lighter component and that of massive component, respectively.}
\label{fig3-9}
\end{figure}

\begin{figure}
\epsfig{figure=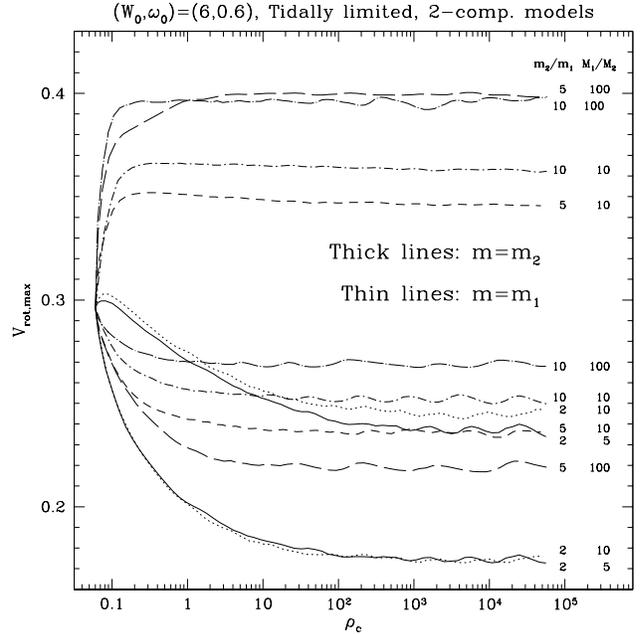, height=0.50\textwidth, width=0.50\textwidth}
\vspace{-3mm}
\caption{Same as the Fig. \ref{fig3-9} but for all two-component models 
with $\omega_0=0.6$ of present study. The $V_{rot,max}$
of the massive stars for the models with higher individual mass ratio 
($m_2/m_1$) even increases very fast during the early evolutionary stage due to the transfer of the
angular momentum from the lighter stars.}
\label{fig3-10}
\end{figure}

Two-body relaxation causes the outward transfer of the angular 
momentum and escape of stars at the outer parts, leading to the decrease 
of rotational energy. Kim et al (2002) showed 
that the maximum value of rotational velocity ($V_{\rm rot,max}$)
in equator decreases monotonically for their single mass models. In 
Fig. \ref{fig3-9}, we displayed the evolution of maximum rotational speed 
at equator, where the effect of the rotation is largest, for
model with the central potential $W_0=6$ and mass function M2A. 
Only four different models for initial degree of rotation is plotted to avoid 
complexity. Different line styles represent the different initial 
rotations. The thick lines represent the run of maximum rotational 
velocity at equator for the high mass component, and the thin lines for 
low mass component. The maximum rotational speed of low mass component 
decreases with the central density. The central density increases with 
time monotonically for core-collapse models as shown in Sect. 3.1. 
The rotational speed for the high mass component, especially for model 
with the initial degree of rotation $\omega_0 = 0.3$ increases during early 
evolutionary stage, and it remains around 0.2 afterward. 
However, the maximum rotational speed for rapidly 
rotating model (i.e., $\omega_0 = 1.6$) decreases with central density 
monotonically for both components. We have shown the run of the
maximum rotational speed at equator for models with the central potential 
$W_0=6$ and $\omega_0=0.6$ in Fig. \ref{fig3-10}. To investigate the 
difference in rate of angular momentum transfer between two mass components,
we include 6 different mass function models. The individual mass ratio 
($m_2/m_1$) and the total mass ratio ($M_1/M_2$) of initial model are 
marked in the figure. 
While the maximum rotation speeds of the high mass component increases 
slightly for model with lowest individual mass ratio ($m_2/m_1=2$), 
it increase more steeply for other models.  The maximum rotational
velocity of high mass component for models with $m_2/m_1 = 5$ and $10$ remains
a nearly constant afterward.

\begin{figure}
\epsfig{figure=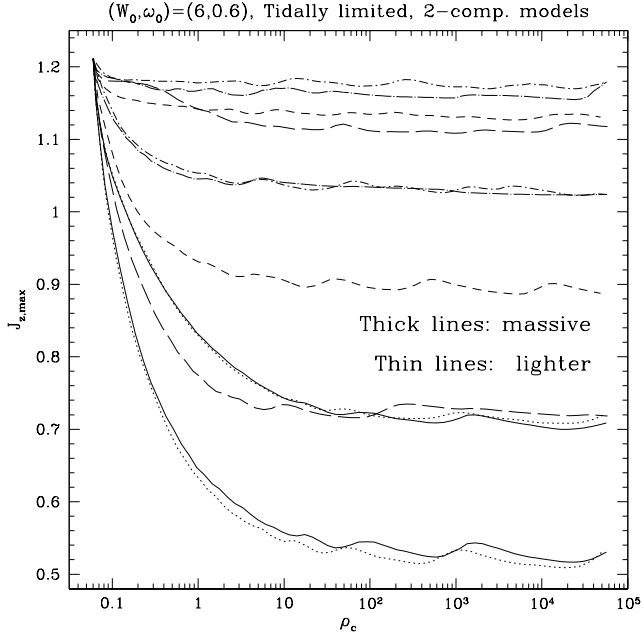, height=0.50\textwidth, width=0.50\textwidth}
\vspace{-3mm}
\caption{Evolution of the maximum specific angular momentum at the equator for the cluster
model with two-component mass functions. The meaning of the line styles are the same to those
of Fig. 3-10.}
\label{fig3-11}
\end{figure}

Kim et al. (2002) found that the specific angular momentum ($J_{z,{\rm max}}$) 
decreases with time for models with central King's potential $W_0=6$. 
Fig. \ref{fig3-11} shows the evolution of maximum specific angular momentum
with the increase of central density for the same models to those of 
Fig. \ref{fig3-10}. 
%
The line style and the thickness of each line represent the same mass 
function model shown in Fig. \ref{fig3-10}. The maximum
angular momentum decreases monotonically irrespective of the mass function 
and the mass component, although the rate of decrease depends 
on the individual mass ratio and the mass components.

\section{Clusters with continuous mass spectrums}

\subsection{ Evolution of central properties}

\begin{figure}
\epsfig{figure=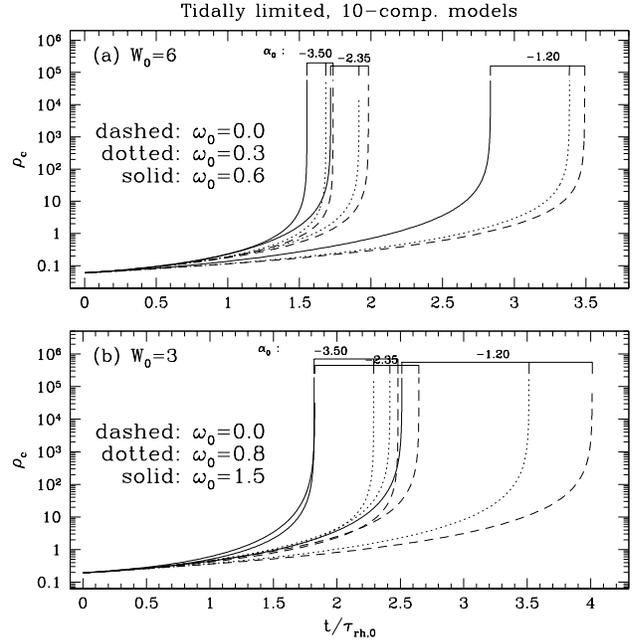, height=0.50\textwidth, width=0.50\textwidth}
\vspace{-3mm}
\caption{Time evolution of the central density for the cluster model with the central concentration
$W_0=6$ (a) and $3$ (b) and continuous mass function (10 mass components). Three different initial
rotation are considered. The cluster models with the same mass function and different initial
rotation is connected with a line. The acceleration of the core-collapse due to the initial
rotation is shown clearly for all models.}
\label{fig3-12}
\end{figure}

Fig. \ref{fig3-12}(a) and (b) shows the time evolution of the total central 
density for cluster models with mass functions MCA, MCB and MCC. 
For each mass function three models with the different initial rotation are
plotted together. 
Since we select the continuous mass function of 
the power-law type and the index of power-law varies with time due to
difference in mass evaporation rates among different mass groups, we 
distinguish the initial power-law index ($\alpha_0$) with the 
power-law index ($\alpha$) at time $t$. The values of the initial 
power-law index are shown in figure and the models with the same mass 
function are connected with lines for clarity. 

Like single component or two-component models, the models with the 
higher initial rotation evolve more rapidly than the those
with lower initial rotation.  Among 
the models with the same amount of the initial rotation the cluster 
with the flatter initial mass function ($\alpha_0 = -1.20$) has a longer 
collapse time, when the time is measured in the initial half-mass 
relaxation time. For models without initial rotation it is known
that the core-collapse time for cluster with steeper mass function is 
shorter than that for the cluster with shallower mass function 
(e.g., Lee, Fahlman \& Richer 1991; Lee \& Goodman 1995; Takahashi 1997).

\begin{figure}
\epsfig{figure=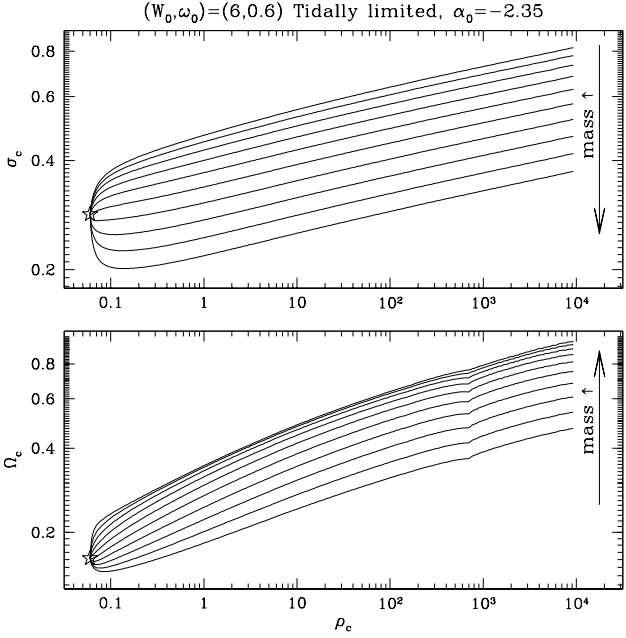, height=0.50\textwidth, width=0.50\textwidth}
\vspace{-3mm}
\caption{Evolution of the 1D central velocity dispersion (a) and the central angular speed (b)
of the individual mass components for a cluster model with $(W_0,\omega_0)=(6,0.6)$ and a
power-law mass function with index $\alpha_0=-2.35$. The staray marks denote the start of evolution.
Not only the velocity dispersion, but the central angular speed shows a self-similar behaviour.}
\label{fig3-13}
\end{figure}

We have shown the evolution of the central one dimensional 
velocity dispersions and the central angular speeds of individual mass 
components for the
with $(W_0, \omega_0) = (6, 0.6)$ and the initial power-law 
slope $\alpha_0=-2.35$ in Fig. \ref{fig3-13}. The evolution starts at 
lower left corner as marked by a starry point in each figure. 
The individual mass of each mass group decreases upward in $\sigma_c$ 
versus $\rho_c$, and vice versa for $\Omega_c$ versus $\rho_c$ as shown 
in Fig. \ref{fig3-13}. The parameters $\gamma$ and $\delta$ which 
characterize the core-collapse phase as defined in Eqs. (10) and (11) are 
$\gamma = 0.103$ and $\delta = 0.113$.

\begin{figure}
\epsfig{figure=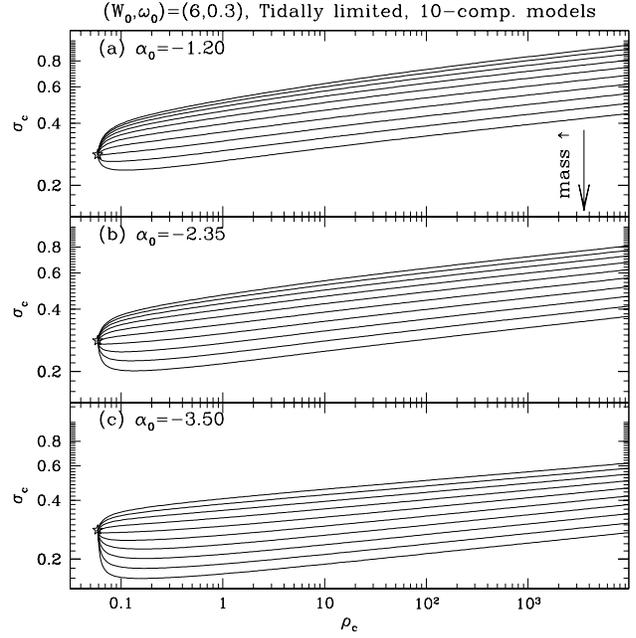, height=0.50\textwidth, width=0.50\textwidth}
\vspace{-3mm}
\caption{Evolution of the 1D central velocity dispersion for cluster models with 
$(W_0,\omega_0)=(6,0.3)$ and a power-law mass function for power-law index (a)
$\alpha_0=-1.20$, (b) $\alpha_0=-2.35$ and (c) $\alpha_0=-3.50$, respectively.
As the slope of mass function increases the central velocity dispersion of the most massive
stars drops faster compared to the other mass function models. The behaviour of $\sigma_c$
on $\rho_c$ shows a self-similarity regardless of the individual mass.}
\label{fig3-14}
\end{figure}

\begin{figure}
\epsfig{figure=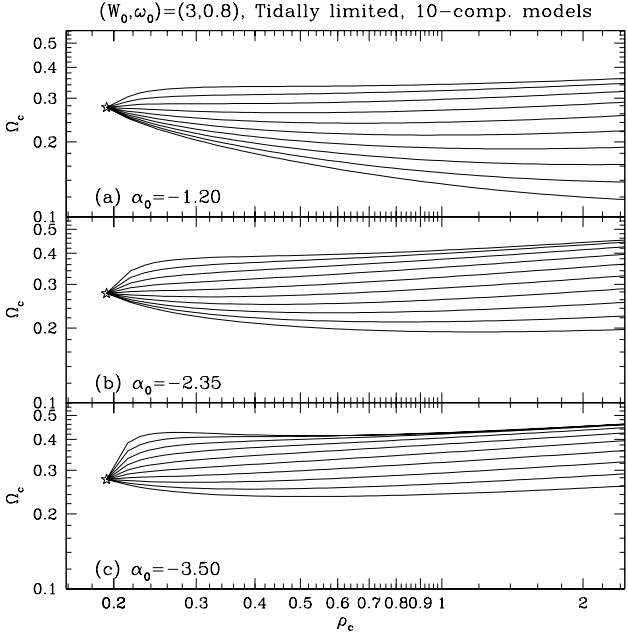, height=0.50\textwidth, width=0.50\textwidth}
\vspace{-3mm}
\caption{Evolution of the central angular speed for cluster models with 
$(W_0,\omega_0)=(3,0.8)$ and a power-law mass function for power-law index (a)
$\alpha_0=-1.20$, (b) $\alpha_0=-2.35$ and (c) $\alpha_0=-3.50$, respectively.
Evolutions during the early stage are shown since the initial rotation affects largely on the
cluster dynamics during the early times.}
\label{fig3-15}
\end{figure}

The effects of the initial mass function on the central velocity dispersion 
and the central angular speed are shown in Figs. \ref{fig3-14} 
\ref{fig3-15}, respectively. We keep the scales of the horizontal axis and 
the vertical axis to be the same for all panels. The direction for
the increase of mass of each mass group is indicated by arrows in the 
figures. For most massive component as $\alpha_0$ decreases $\sigma_c$ 
decreases more rapidly, driving more rapid evolution of the cluster as 
indicated by earlier core-collapse. The development of the mass segregation 
is established more quickly for the model with the steeper mass function. 
For $\Omega_c$ versus $\rho_c$ (Fig. \ref{fig3-15}) we have shown only 
the early stage of evolution since the rotation affects mainly during 
the initial period. It is evident that the development of 
self-similar core-collapse phase occurs earlier for the model with the 
steeper mass function than that with flatter mass function. 
The central angular speed at a given central density is larger for the 
model with the steeper mass function.
The rapid increase of the central 
angular speed for the highest mass group and the slow 
decrease for the lowest mass group for the model
with $\alpha_0 = -3.50$ is a consequence of the rapid development of mass 
segregation. The steep decrease of central velocity dispersion drives the 
increase of the central angular speed because of negative specific 
moment of inertia. 

\subsection{Rotational properties}

\begin{figure}
\epsfig{figure=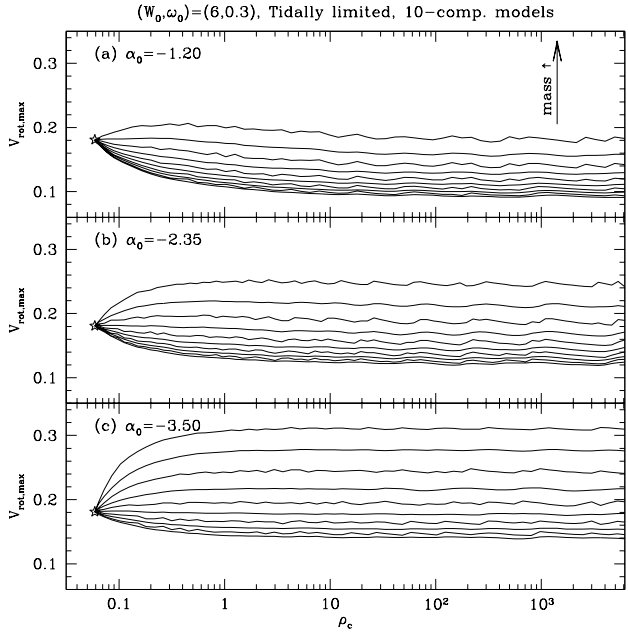, height=0.50\textwidth, width=0.5\textwidth}
\vspace{-3mm}
\caption{Evolution of the maximum rotational velocity at equator for cluster models with 
$(W_0,\omega_0)=(6,0.3)$ and a power-law mass function for power-law index (a)
$\alpha_0=-1.20$, (b) $\alpha_0=-2.35$ and (c) $\alpha_0=-3.50$, respectively.
For the steepest mass function ($\alpha_0=-3.50$), $V_{rot,max}$ of the massive stars increase much
than that of the other models with slowly increasing mass functions.} 
\label{fig3-17}
\end{figure}

\begin{figure}
\epsfig{figure=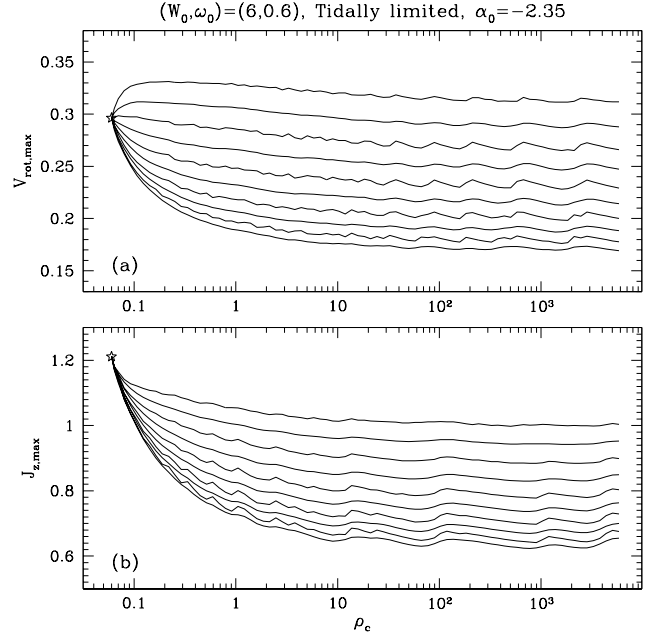, height=0.50\textwidth, width=0.50\textwidth}
\vspace{-3mm}
\caption{Evolutions of (a) the maximum rotational velocity ($V_{rot,max}$) and (b) the maximum
specific angular momentum ($J_{z,max}$) at the equator for a model with 
$(W_0,\omega_0,\alpha_0)=(6,0.6,-2.35)$. While $V_{rot,max}$ of the massive stars increases with
time (i.e., increasing $\rho_c$), $J_{z,max}$ decreases monotonically due to the outward
transfer of the angular momentum.}
\label{fig3-16}
\end{figure}

Fig. \ref{fig3-16} displays the evolution of the maximum rotational 
velocities of individual mass components at the equator for the model with 
$(W_0,\omega_0) = (6,0.3)$. We included the models with mass functions 
MCA, MCB and MCC together. The total mass of each mass group increases along 
the arrow shown in Fig. \ref{fig3-16}(a). The evolution of
rotational speed for the cluster with the continuous mass function is 
similar to that of the two component models. For the model  with 
$\alpha_0 = -3.50$ the maximum rotational speed of the mass component of 
$m_5$ remains nearly a constant. Since the radius where the rotational velocity
becomes maximum at the equator decreases with the time, the mass 
group with $m_5$ rotates faster with time.The constant behaviour of the 
maximum rotational speed occurs for the mass group of $m_7$ for model 
with $\alpha_0 = -2.35$ and mass group $m_9$ for model with 
$\alpha_0 = -1.20$, respectively. Fig. \ref{fig3-17}(a) and (b) show 
the evolution of the maximum rotational speed and the evolution of the 
maximum angular momentum at equator for model with $(W_0,\omega_0) = (6,0.6)$ 
and mass function with $\alpha_0 = -2.35$. The maximum rotational speed of the 
lower mass groups decreases more than the cluster rotating slowly 
(see Fig. \ref{fig3-16}(b) for comparison). While the high mass component 
rotates faster, the angular momentum decreases continuously through 
whole evolutionary phase. Since the outward transfer of angular momentum 
for each mass species is closely related with the two-body
relaxation process, the evolution of the maximum of the angular momentum 
reaches a plateau when it is measured in the central density.

\section{Evolution after core-collapse}
So far we have concentrated our discussion until the core-collapse. We now
discuss the evolution beyond the core collapse.

\subsection{Central density and mass loss}

\begin{figure}
\epsfig{figure=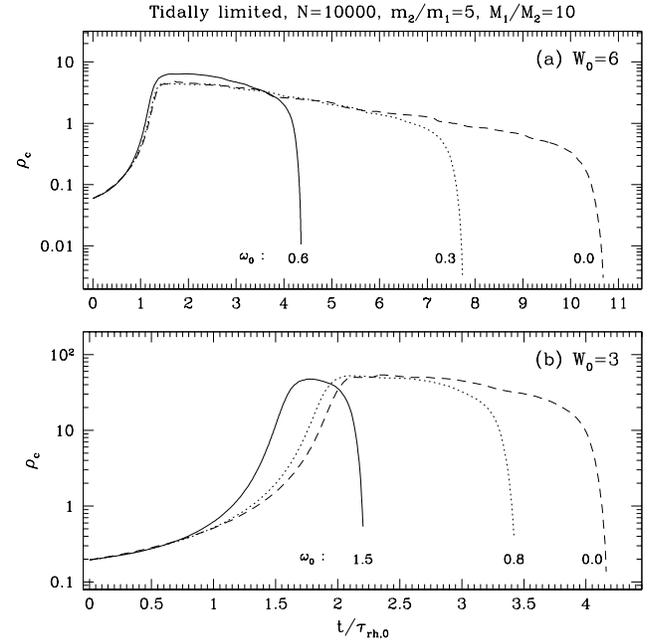, height=0.50\textwidth, width=0.5\textwidth}
\vspace{-3mm}
\caption{Evolution of the central density for the cluster models with the central potential
(a) $W_0=6$ and (b) $W_0=3$ for mass function $(m_2/m_1,M_1/M_2)=(5,10)$. The initial number
of stars in a cluster is $N=10000$ for different initial rotational parameters
as indicated in the figure.
}
\label{fig3-18}
\end{figure}

\begin{figure}
\epsfig{figure=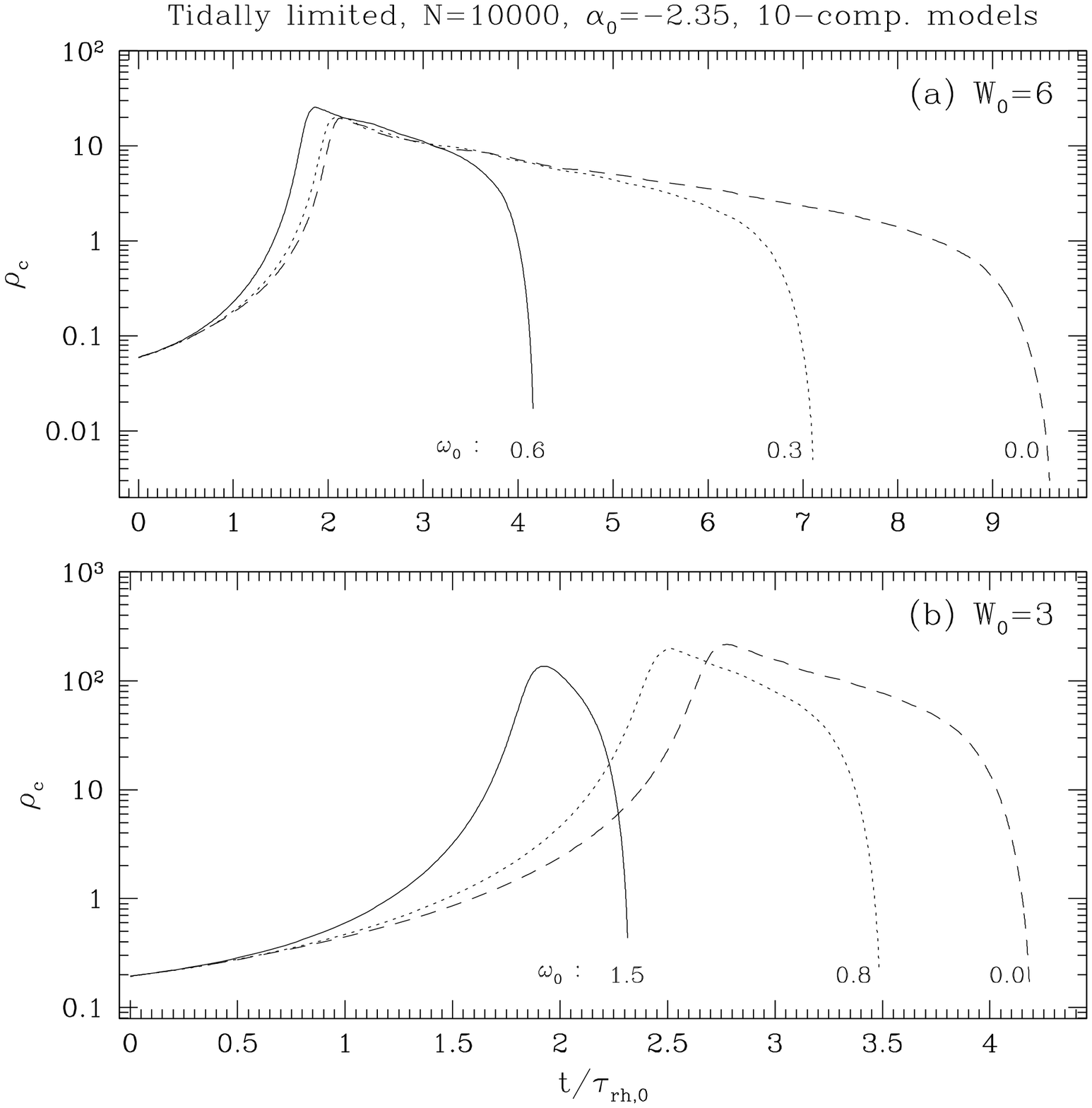, height=0.50\textwidth, width=0.50\textwidth}
\vspace{-3mm}
\caption{Same as the Fig. \ref{fig3-18} but for models with 
power-law mass function 
($\alpha_0=-2.35$).}
\label{fig3-19}
\end{figure}

Figs. \ref{fig3-18} and \ref{fig3-19} show the time evolution of the central 
density of the rotating stellar systems beyond the core collapse. 
The evolutions for two component models with the central potential 
$W_0 = 6$ (Fig. \ref{fig3-18}(a)) and 3 (Fig. \ref{fig3-18}(b)) are 
displayed for clusters with three different initial rotations 
($\omega_0 = 0.0, 0.3$ and $0.6$ for $W_0 = 6$, and $\omega_0 = 0.0, 0.8$ 
and $1.5$ for $W_0 = 3$). For two component clusters we
employed the mass function M2C, while the mass function MCB for the 
continuous mass spectrum is included. The models with the largest initial 
rotation (solid lines) reach the collapse earlier than the  
models with smaller initial rotation (dashed and dotted lines).
We can clearly see the acceleration of the evolution during pre and post
collapse phases due to initial rotation.

\begin{figure}
\epsfig{figure=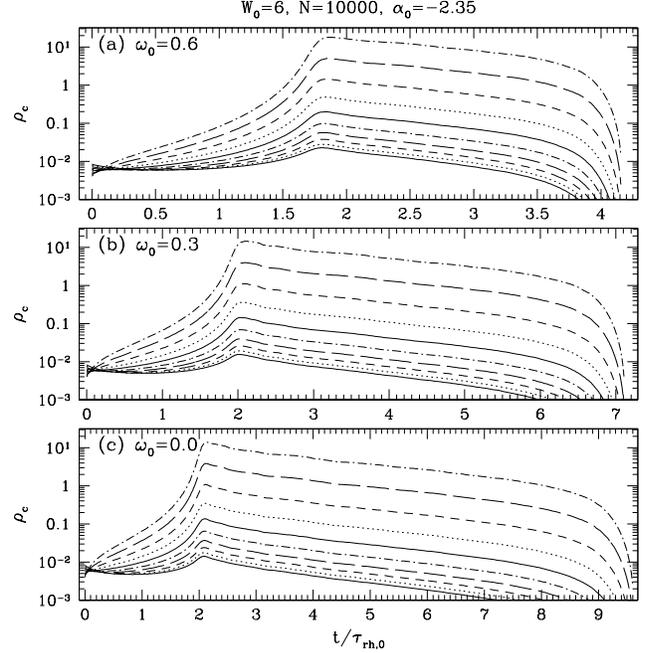, height=0.50\textwidth, width=0.50\textwidth}
\vspace{-3mm}
\caption{Time evolution of the central density of the individual mass component for models
with $(W_0,\alpha_0,N)=(6,-2.35,10000)$ and the initial rotations of (a) $\omega_0=0.6$,
(b) $\omega_0=0.3$ and (c) $\omega_0=0.0$, respectively. $\rho_c$ of the massive stars
increases very quickly due to the development of the mass segregation.}
\label{fig3-20}
\end{figure}

Fig. \ref{fig3-20} displays the run of the individual central density for 
models with the continuous mass spectrum. The mass segregation due to 
the energy equipartition during the early stage is shown clearly for all the
present models. The run of the central density after core bounce roughly 
follows a power-law of $\rho_c \propto t^{\beta}$, 
for all mass groups. The index of power-law tends to be smaller for
lower mass components.  For the non-rotating model 
we find $\beta = -1.54$ for the
highest mass group ($m_{10}$) and $\beta = -1.88$ for the lowest mass 
group ($m_1$). These values are not far from the power law 
slope $\beta = -2$ of the central density with time which was derived by 
Kim, Lee \& Goodman (1998), for two-component 
models. These slopes increase with 
the initial rotation: $\beta = -1.83$ and $-2.42$ for the 
$m_{10}$ and the $m_1$, respectively, for model with $\omega_0 = 0.6$. 

\begin{figure}
\epsfig{figure=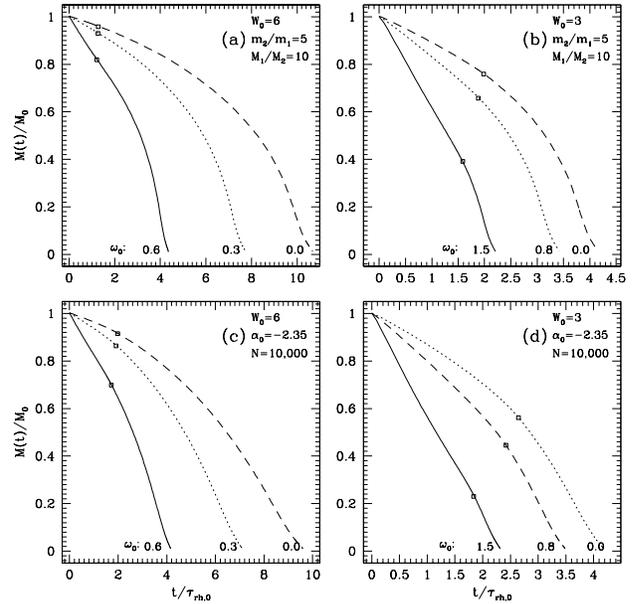, height=0.50\textwidth, width=0.50\textwidth}
\vspace{-3mm}
\caption{Evolution of the total mass retained in a cluster. The parameters characterize the
cluster model are written on the upper-right corner of the each panel. The epochs of
core-collapse are marked with the open squares.}
\label{fig3-21}
\end{figure}

We have shown the evolution 
of the total mass of the cluster for all post-collapse models in 
Fig. \ref{fig3-21}. The initial mass function, the central King's potential 
and the degree of rotation are shown in each panel. The epochs of
core-collapse are marked by open squares. Note the marked 
core-collapse time is obtained not from the present post-collapse models but 
from the pre-collapse models discussed in \S 3. 
Fig. \ref{fig3-21}(a) and (b) are the results 
obtained from two component models. The total mass of rotating cluster
decrease more rapidly than the non-rotating cluster not only pre-collapse 
phase, but also after core-collapse. The raid decrease of total mass for 
rotating model results in  smaller tidal radius.

\subsection{Velocity dispersion and angular speed}

\begin{figure}
\epsfig{figure=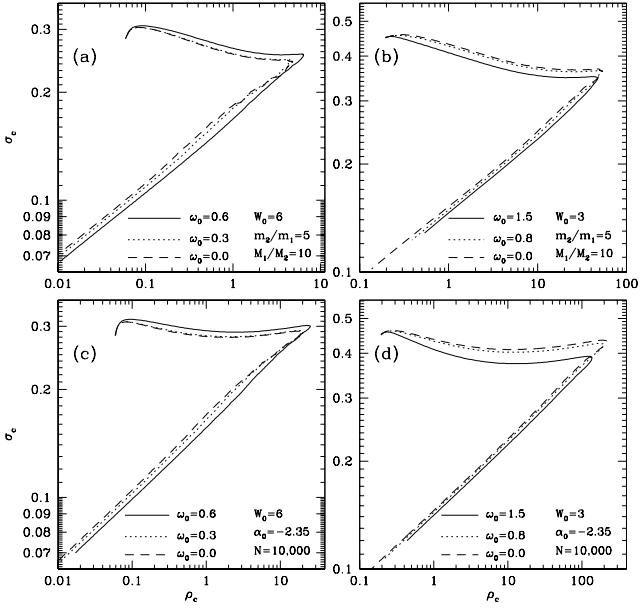, height=0.50\textwidth, width=0.50\textwidth}
\vspace{-3mm}
\caption{Evolution of the mass-weighted 1D central velocity dispersion on the central
density. The model clusters shown here are the same to the models displayed in Fig. 3-21.
The behaviour of $\sigma_c$ on $\rho_c$ after core-bounce shows a power-law signature
irrespective of the amount of the initial rotation.}
\label{fig3-22}
\end{figure}

The evolution of the mass-weighted central velocity dispersion ($\sigma_c$)
as a function of the central density ($\rho_c$) is shown in 
Fig. \ref{fig3-22}. 
The evolution of the central velocity dispersion after 
core-bounce for all post-collapse models shows a similar trend, 
i.e., the effect of 
rotation on the central velocity dispersion after core-bounce is very small. 
Since the core rotates like a 
rigid body, the rotation velocity near the cluster center is 
negligible. While the run of $\sigma_c$ on $\rho_c$ during pre-collapse shows 
the power-law behaviour due to self-similarity of collapsing 
core for the single mass system (Cohn 1980, Kim et al 2002), the stellar 
system with mass spectrum does not show a simple power-law in the early
phase as explained in Sect. 3.2, due to the mass segregation. 
The behaviour of $\sigma_c$ on $\rho_c$ after 
core-bounce can be, however approximated as a simple power-law even for 
the multi-mass systems because the mass-segregation was already
established at the time of the beginning of the expansion. 
The power-law index $\gamma$ can be predicted by applying energy 
balance argument (Kim et al. 2002). Inspection of Fig. \ref{fig3-22} shows 
$\gamma \sim 0.42$, slightly smaller than the power-law index of $\sim 0.33$
obtained for the single mass system.

\begin{figure}
\epsfig{figure=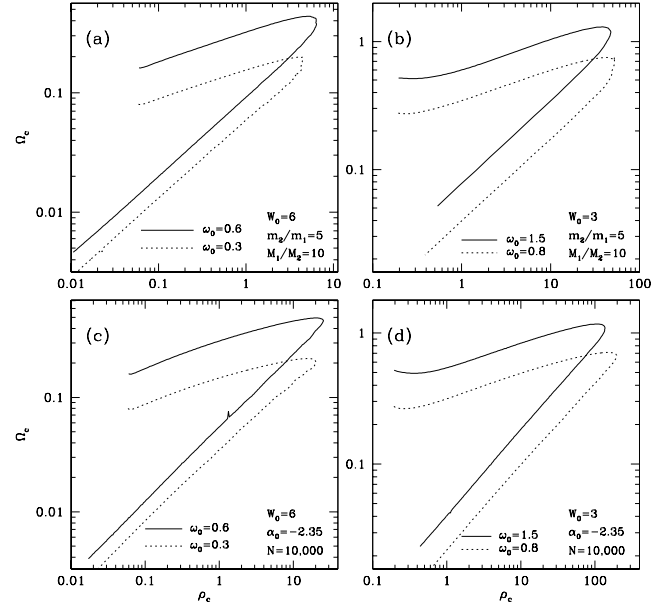, height=0.50\textwidth, width=0.50\textwidth}
\vspace{-3mm}
\caption{Same as the Fig. \ref{fig3-22}, 
but for mass-weighted central angular speed on the central
density. Only rotating models are shown here. A small spike shown in figure (c) is mainly
due to the numerical error.}
\label{fig3-23}
\end{figure}

\begin{figure}
\epsfig{figure=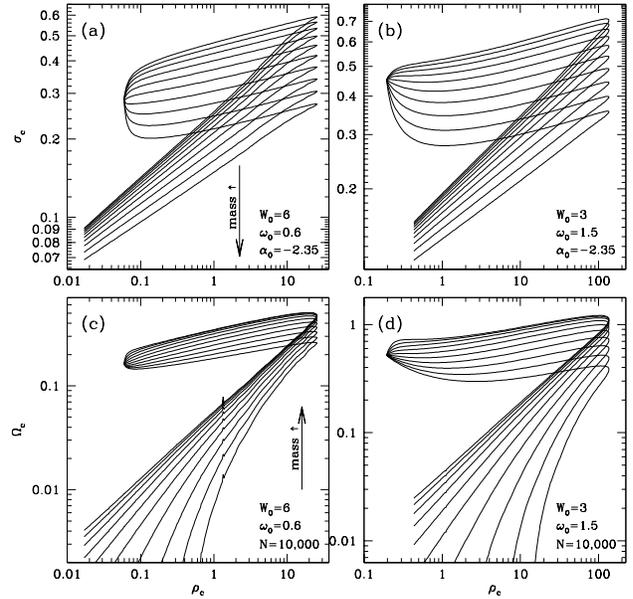, height=0.50\textwidth, width=0.50\textwidth}
\vspace{-3mm}
\caption{Evolution of the central velocity dispersion (a,b)and the central angular speed (c,d)
of the individual mass component for model clusters with the fastest initial rotation
($\omega_0=0.6$ for $W_0=6$ and $\omega_0=1.5$ for $W_0=3$). In $\sigma_c$ on $\rho_c$ plots,
the individual mass component show a power-law behaviour, though the index of the power-law
depends on the mass of the individual star. The direction increasing individual mass is denoted
with arrows in panel (a) and (c), respectively.}
\label{fig3-24}
\end{figure}

In Fig. \ref{fig3-23}, we have shown the run of the mass-weighted central 
angular rotational speed ($\Omega_c$) on the central density ($\rho_c$) 
for the whole post-collapse models which have the initial rotation. The
angular speed also appears to follow roughly power law on $\rho_c$ during 
both pre- and post-collapse, except for the early evolutionary stage in the 
pre-collapse phase for the models with $W_0 = 3$. 
The evolutions of $\Omega_c$ on $\rho_c$ after the very early stage shows 
a similar trend irrespective of the degree of the initial rotation. 
It implies that the effect of the initial rotation for the cluster center 
disappears quickly. The evolutions of $\sigma_c$ and $\Omega_c$ on 
$\rho_c$ of the individual mass component for models with the continuous 
mass spectrum is displayed in Fig. \ref{fig3-24}. The run of $\sigma_c$ is 
represented well by a simple power-law. The higher the stellar mass,  the 
shallower the power-law index. While the mass-weighted central angular 
rotation speed appears to show the power-law behaviour on $\rho_c$,  
$\Omega_c$ of the low mass stars deviates from the power-law type 
evolutionary behaviour during post-collapse phase. It shows much steeper 
decrease than higher mass components, though the 
contribution of the lower mass component on the total $\Omega_c$ is 
negligible.

\subsection{Rotational profiles}
\begin{figure}
\epsfig{figure=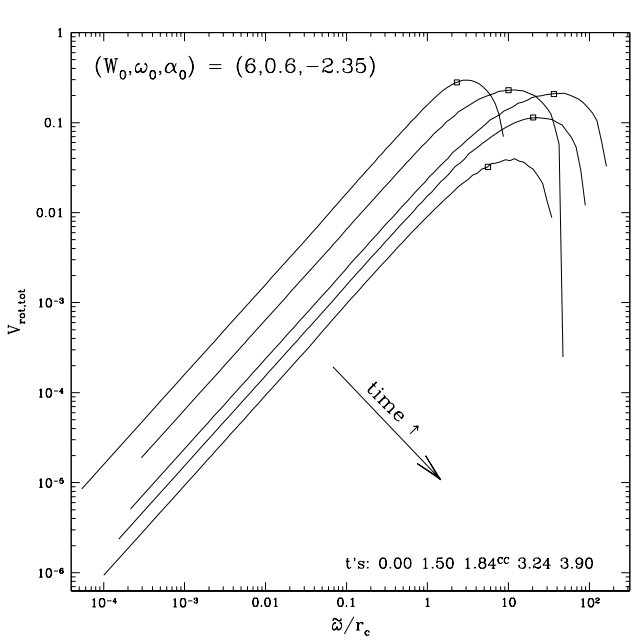, height=0.50\textwidth, width=0.50\textwidth}
\vspace{-3mm}
\caption{Radial profiles of the rotational velocities at a few selected epochs for a 
cluster model with $(W_0,\omega_0,\alpha_0)=(6,0.6,-2.35)$. Half-mass radii ($r_h$)
at corresponding
epochs are marked with open squares. At the initial time, the rotational inside $r_h$
follows a rigid body rotation. Note that the radii is measured with the current
core radii.}
\label{fig3-25}
\end{figure}

\begin{figure}
\epsfig{figure=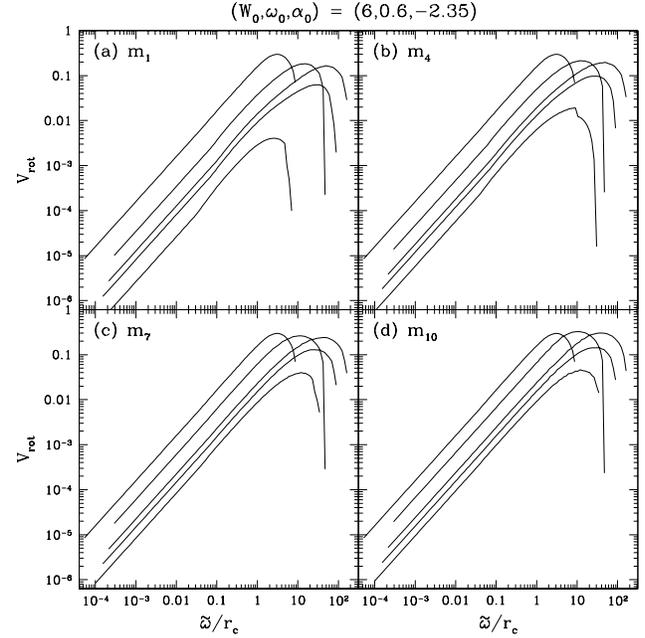, height=0.50\textwidth, width=0.50\textwidth}
\vspace{-3mm}
\caption{Same as the Fig. \ref{fig3-25} but for a few selected 
mass components, (a) the lowest mass
component $m_1$, (b) $m_4$, (c) $m_7$ and (d) $m_{10}$, the most massive star. 
The rotational structure of the massive stars extends 
farther than that of the less massive stars.}
\label{fig3-26}
\end{figure}

In Figs. \ref{fig3-25} and \ref{fig3-26}, we have shown the radial profiles 
of the rotational velocity at equator for model with 
$(W_0,\omega_0) = (6,0.6)$ and continuous mass spectrum with 
$\alpha_0 = -2.35$. The run of the rotational velocity for the selected 
mass components ($m_1, m_4, m_7, m_{10}$) is displayed in Fig. \ref{fig3-26}.
Epochs shown in this plot are 
$t/\tau_{rh,0} = 0.00, 1.50, 1.84, 3.24$ and $3.90$, where 
the core-collapse occurs at $t/\tau_{rh,0} = 1.84$. At $t/\tau_{rh,0} = 3.90$ 
the cluster is nearly completely dissolved. 
The half-mass radii at these each evolutionary times 
are marked with open squares and the direction with increasing times is 
shown with arrow in Fig. \ref{fig3-25}. Note that x-axis, i.e., radius of 
cylindrical coordinate is measured in units of the current core radius.

Total rotational velocity decreases continuously with time due to 
the loss of the angular momentum through the cluster boundary. 
The outward transfer of the angular momentum causes the increase of
the radius when it is measured in units of current core radius, 
where the rotational velocity 
becomes the maximum. The radius of maximum rotational velocity decreases 
after core bounce due to the higher mass-loss. While the changes of 
angular momentum due to the outward transfer and the loss through the 
cluster boundary changes the structure of the velocity profile beyond the 
half-mass radii, the shape of rotational velocity inside the cluster core 
remains to be rigid body rotation.

\subsection{Evolution of mass function}

\begin{figure}
\epsfig{figure=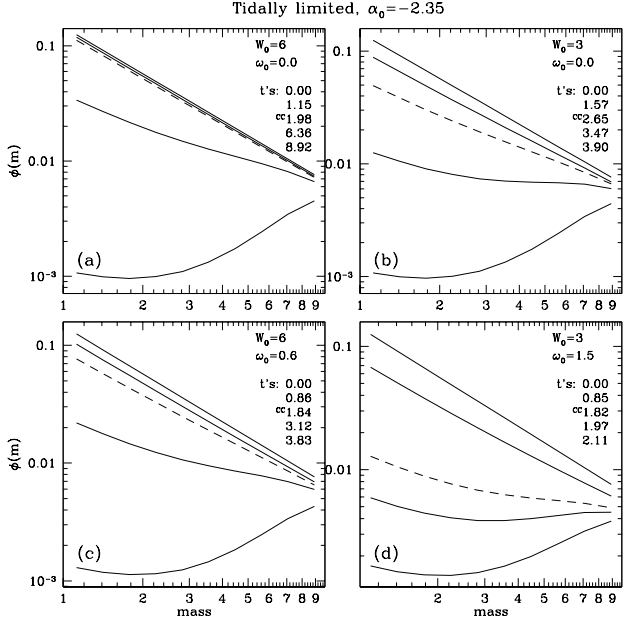, height=0.50\textwidth, width=0.50\textwidth}
\vspace{-3mm}
\caption{Mass functions of the several selected epochs for the cluster 
models with the continuous initial mass function with a power-law index 
$\alpha_0=-2.35$. For a comparison clusters without the initial rotation 
(Figs. a and b) and the clusters with
the highest rotation (Figs. c and d) for a given central potential are 
displayed. The epochs
where the core-collapse happened are distinguished with dashed lines from 
the other evolutionary
epochs. The change of mass function measured by the power-law index as a 
function of the cluster mass in units of initial mass is shown in next
figure Fig. \ref{fig3-28}).}
\label{fig3-27}
\end{figure}

\begin{figure}
\epsfig{figure=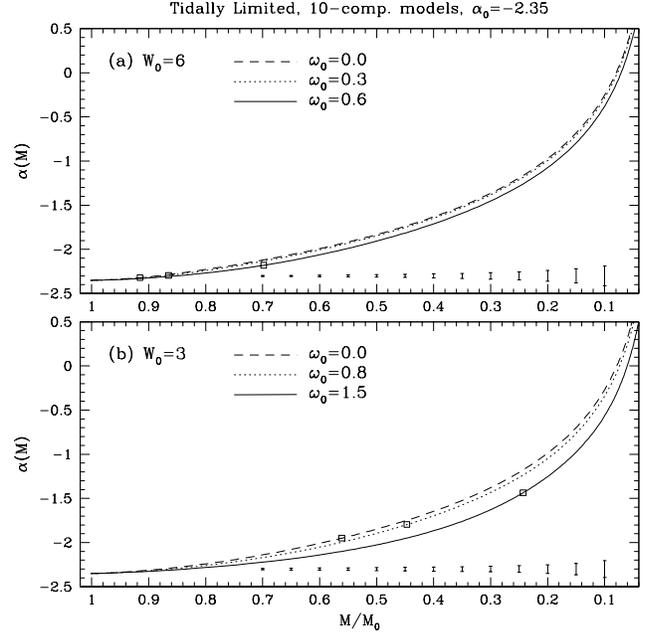, height=0.50\textwidth, width=0.50\textwidth}
\vspace{-3mm}
\caption{Evolution of the power-law index $\alpha_M$ on the total 
cluster mass for whole
post-collapse models with a continuous mass function. Open squares 
represent the core
collapse times for each cluster model. The errorbar on 
each panel represent the typical
errorbar of the fastest rotating model when performing a linear 
least square fitting for
a double logarithm plot such as the panels in Fig. \ref{fig3-27}. 
The slope of the power-law mass
function for a cluster with the fastest initial rotation is slightly lower than that of
non-rotating model.}
\label{fig3-28}
\end{figure}

Mass segregation drives the concentration of the higher mass stars 
in the cluster center while 
the lower mass stars go to the outer halo of the cluster. This leads to the 
preferential evaporation of lower mass stars through the the tidal boundary. 
Therefore, low mass stars are selectively depleted 
from the initial mass function, resulting a 
change of the mass function with time. 
Near the end of the evolution,
the mass function can even be inverted, i.e., there are more high mass stars
than lower mass stars. 
The global mass function $\phi(m)$ 
at several selected epochs for four different models 
with the continuous mass spectrum is displayed in Fig. \ref{fig3-27}. 
We include the non-rotating models and the rapidly rotating models in order 
to study the effect of the initial rotation on the mass function. The shape
of the mass function at the time of core-collapse is distinguished by dashed 
line from the mass function at the other epochs. The evolution of power-law 
index ($\alpha(M)$) for whole post-collapse models with the continuous mass 
spectrum is shown is Fig. \ref{fig3-28}. 
The mass function deviates from the power-law as a result of
mass-dependent evaporation rates. 
In Fig. \ref{fig3-28}, 
we have shown the typical error of the fitted power-law index for the
models which has the highest initial rotation. The epochs where the 
core-collapse occurs is marked with open squares. In the figures,
we have chosen $M/M_0$ for the abscissa instead of time 
because the life-time of clusters are significantly different for
different models.

It is evident that the behavior of
$\alpha$ does not depend sensitively on the degree of rotation, although
the rotation causes less change of $\alpha$.
Although the effect is small, the rotation causes slower changes in
$\alpha$ as a function of $M/M_0$. The angular momentum exchange among
different mass species appears to act against the mass segregation.

\begin{figure}
\epsfig{figure=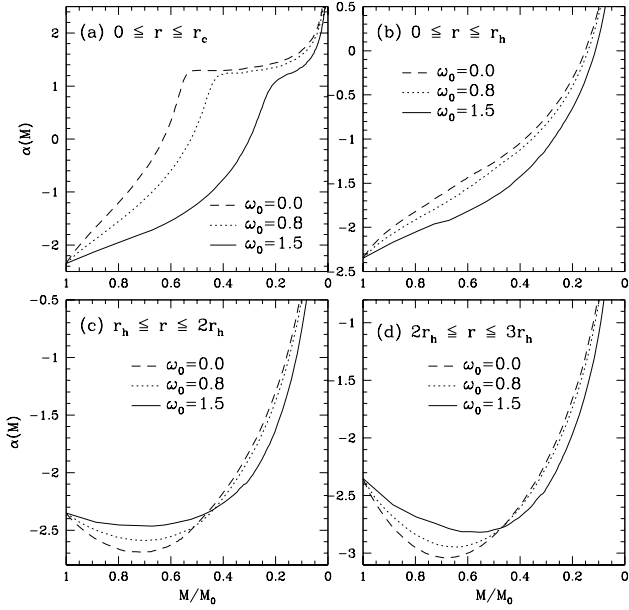, height=0.50\textwidth, width=0.50\textwidth}
\vspace{-3mm}
\caption{Same as the Fig. \ref{fig3-28}
but for four different radial regions of cluster models
with the central concentration $W_0=3$, (a) $r < r_c,\:$
(b) $r < r_h,\:$, (c) $r_h < r < 2r_h$ and (d) $2r_h < r < 3r_h\,$, 
respectively.  A combined effect due to the mass segregation and 
the initial rotation is shown clearly in panels (c) and (d).}
\label{fig3-29}
\end{figure}

The mass function also becomes dependent on the location
as the mass segregation proceeds. The existence of initial rotation and 
its outward transfer also affect the energy exchange process that causes
mass segregation. In order to see the role of rotation on the mass segregation,
we have compared the evolution of $\alpha$ at four different
locations in Fig. \ref{fig3-29}. The different locations are chosen such that
$r < r_c,\: r < r_h,\: r_h < r < 2r_h$ and $2r_h < r < 3r_h$.
The slope of mass 
function inside the core (Fig \ref{fig3-29}a) increases with time 
during pre-collapse
evolution. At the time of the core-collapse the shape of the mass spectrum 
become inverted for all models. The power-law index $\alpha$ at this 
evolutionary stage remains $\sim 1.3$ with a weak dependence on 
the the degree of the 
initial rotation. The mass function within 
the half-mass radius is flattened continuously with the decrease of total mass. 
The model without the initial rotation has a rather shallower mass 
function at fixed total mass compared to that of the cluster
rotating fast. Initial rotation plays an important role in removing the 
stars through the tidal boundary including both the lower and higher mass 
stars. 
Compared to non-rotating models, high mass stars
in rotating models appear to have higher chance of evaporation.
Thus the mass function changes more slowly for
rotating models than non-rotating models. 
Since the effect of the initial rotation remains in region 
$r_h < r < 2r_h$, it is expected that the shapes of mass function in this 
region shows significant dependence on rotation. The concentration of
the high mass component in the cluster center drives the steepening of the 
mass function near the half-mass radius. When the development of 
mass segregation is settled in, the slope of the mass function decreases
with time due to the selective evaporation of the lower mass stars through 
the tidal radius. The evolution of $\alpha$ for the highly rotating model 
shows a different profile compared to that of non-rotating model
at this radial region. While there is a central concentration of the 
massive stars, the angular momentum which has the maximum around $r_h$ 
pushes the high mass stars outward. Therefore, there is only a little
change in $\alpha$ during the early evolutionary stage for
rapidly rotating models.

\section{Conclusion and Discussion}
We have studied the dynamical evolution of the rotating stellar systems 
with the mass spectrum by solving the orbit-averaged 2D FP equation in 
($E,J_z$) space. Numerical simulations are performed both for simple 
two component clusters and for clusters with a power-law mass function
represented by ten mass species. 
In order to explore the evolution 
after core collapse we add the heating by three-body processes. We have
employed rotating King models as initial models, where the 
velocity dispersions for all mass components are equal, i.e., no mass 
segregation at the beginning. The rotating King models are characterized
by two parameters: initial central potential ($W_0$) and degree of 
initial rotation ($\omega_0$). Clusters with two different central potential 
$W_0 = 6$ and $3$ are studied extensively. For models only until 
core-collapse, we have studied the evolution of the clusters with 
six different mass 
functions for two-component models, while three models are studied for 
clusters with a continuous mass function (power-law). For models where the
evolution beyond the core-collapse are explored, we consider clusters 
with a mass function M2C ($m_2/m_1 = 5, M_1/M_2 = 10$) for
the two component model and clusters with $\alpha_0 = -2.35$ for the model 
with continuous mass spectrum.

Our results show that, as in equal mass system, the presence of the 
initial rotation accelerates the dynamical evolution as manifested by
rapid core collapse and dissolution. 
The degree of the acceleration depends 
on the amount of the initial rotation and the shape of mass function. 
As the ratio of the mass ($m_2/m_1$) increases the degree of the 
acceleration decreases for 
two component models since both mass segregation and the rotation compete
in acceleration of core-collapse. If $m_2/m_1$ is very large, the
mass segregation alone could significantly reduce the core-collapse
time and there is not much room for rotation to accelerate further.
For models with the power-law mass function, the 
acceleration rate of core-collapse due to the rotation is larger for 
the model with a steeper slope of the mass function for a given  
initial rotation parameter ($\omega_0$). The shortening of the life-time 
(dissolution of cluster) due to the rotation is observed far beyond the 
core bounce. The increase of mass loss rate, resulting from the enhanced 
two-body relaxation process causes the faster dissolution of cluster.

The evolution of $\sigma_c$ on $\rho_c$ can be approximated by a 
power-law except for the early evolutionary stage regardless of the degree 
of rotation both for pre- and post-collapse. The measured power law 
index $\gamma$ for pre-collapse is very close to the value obtained for 
single mass system, while we have obtained a slightly shallower slope for 
post-collapse phase. The development of mass segregation, 
a consequence of the 
evolution of the multi-mass system is visible clearly both for 
non-rotating and rotating models. The evolution of $\Omega_c$ on 
$\rho_c$ shows a power-law behaviour, too. The slope of the power-law is,
however, larger than that obtained for the single mass cluster. While the
angular momentum is transferred only outward for the equal mass system, 
the exchange of the angular momentum between different mass species 
occurs for the multi-mass system, resulting a faster increase
of $\Omega_c$ on $\rho_c$. Due to a cooperation of the central concentration 
of the massive stars (mass segregation) and the transfer of angular momentum 
from high mass to low mass stars, the radii where $V_{rot}$ 
reaches the maximum value goes outward. The maximum rotation of the most 
massive stars even increases at the early times, while it shows a monotonic 
decrease for the single mass system.

\begin{figure}
\epsfig{figure=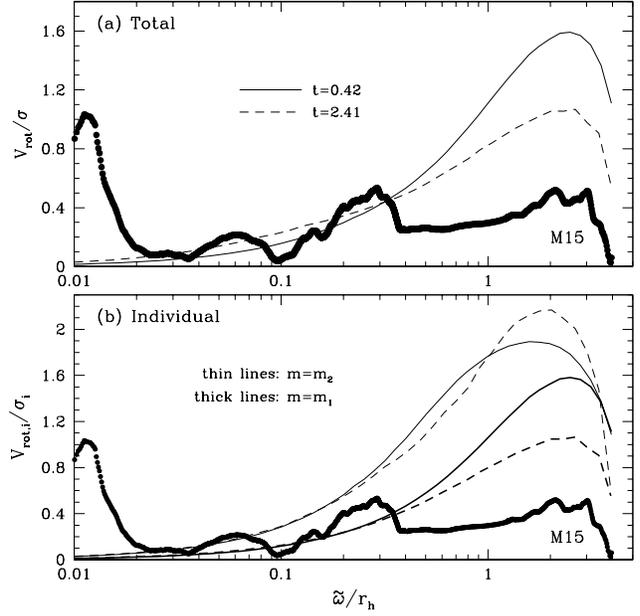, height=0.50\textwidth, width=0.50\textwidth}
\vspace{-3mm}
\caption{Comparison of observed radial profile of the rotational 
velocity over 1D velocity dispersion
for the galactic globular cluster M15 (data provided by Gebhardt 2002) with
the result of a selected two-component model with
$(W_0,\omega_0)=(6,0.6)$, $(m_2/ m_1, M_1/M_2)=(5,10)$.}
\label{fig3-30}
\end{figure}

There are a few observations regarding the direct measurement of the 
rotation parameter ($V_{rot}$) (Gebhardt et al. 1995). Recently Gerssen 
et al. (2002) reported the kinematical study of the central
part of the globular cluster M15, including $V_{rot}$. M15 is known as the 
globular cluster which contains a collapsed core. The radial profile of 
$V_{rot}/\sigma$ (rotational velocity over one dimensional velocity 
dispersion) shows a steep increase and a rapid decline followed by
a slow rise. The steep rise of $V_{rot}/\sigma$ 
near the cluster center can not be explained by a single mass model 
(Kim et al. 2002). 
We have shown the run of 
$V_{rot}/\sigma$ 
for two component model in Fig. \ref{fig3-30} at two selected epochs, 
one in pre-collapse ($t/\tau_{rh,0} = 0.42$) and the other after core bounce 
($t/\tau_{rh,0} = 2.41$), together with the observed data by Gebhardt (2002). 
The radius is measured in current 
half-mass radius. It is evident that cluster loose the angular momentum 
through the tidal boundary. The behaviour of the mass-weighted 
$V_{rot}/\sigma$ (Fig. \ref{fig3-30}(a)) are roughly similar to that of 
observed profile of M15 except for the central region. 
Note that the known 
half-mass radius of M15 is $\sim 3^{'}.09$ (Djorgovski 1993). 
For the high mass component, $V_{rot,i}/\sigma_i$ 
where $i$ represents the individual mass component, at $t = 2.41$ is 
higher than that obtained at $t = 0.42$, especially beyond $r_h$. It is 
mainly due to the lower velocity dispersion after core bounce, not 
due to the higher value of the rotation velocities. The highly rotating 
central region of M15 is, as in single mass system, not explained with the 
current multi-mass models, although there is difference in rotational 
structure between the single mass and the multi-mass systems. 
From the kinematical study of the central region of M15, 
Gerssen et al. (2002) 
proposed the presence of intermediate mass black hole (IMBH) with 
$M = 3.9\times10^3 M_{\odot}$. On the contrary, Baumgardt et al. (2002) 
demonstrated that the rise of velocity dispersion into the center can be 
explained with the clustering of the remnant stars, which are expected to be 
overwhelmingly populated in the cluster core than the normal
stars. However, they did not rule out the possibility of the presence of IMBH. 
It may be necessary to include the remnant stars and the IMBH in current 
2DFP models to investigate the role of the initial rotation on the observed 
strong increasing of $V_{rot}/\sigma$ near the center of M15.

The models presented here still neglect many important physical processes
occurring in real star clusters. The stellar evolution and primordial binaries
are known to affect the early evolution of globular clusters. 
The possible existence of intermediate mass black hole in the center
could also affect the course of dynamical evolution. These will be
the task of future works.

This work was supported by the Korea Research Foundation Grant No. D00268 
in 2001 to HML and by SFB439 to RS.


\begin{thebibliography}{}

\bibitem{} Anderson, J. \& King, I.R., 2003, ApJ, 126, 772
\bibitem{}Baumgardt, H., Hut, P., Makino, J., McMillan, S.L.W., 
\& Portegies Zwart, S., 2002, ApJ, 582, L21
\bibitem{}Cohn H., 1979, ApJ, 234, 1036
\bibitem{}Cohn H., 1980, ApJ, 242, 765
\bibitem{}Djorgovski, S. G., 1993, in  Structure and Dynamics of Globular
Clusters, eds. S. G. Djorgovski and G. Meylan, Astronomical Society of Pacific,
p373
\bibitem{}Einsel C., Spurzem R., 1999, MNRAS, 302, 81
\bibitem{}Fregeau, J.M., Joshi, K.J., Portegies Zwart, S.F., \& Rasio, F.A., ApJ, 2002, 570, 171
\bibitem{} Gao, B., Goodman, J., Cohn, H., \& Murphy, B., 1991, ApJ, 370, 567
\bibitem{} Gebhardt, K., 2002, private communication
\bibitem{} Gebhardt, K., Pryor, C., Williams, T. B., \& Hesser, J. E., 1995, AJ,
110, 1699
\bibitem{} Gerssen, J., van der Marel, R. P., Gebhardt, K., Guhathakurta, P.,
Peterson, R., \& Pryor, C., 2003, ApJ, in press.
\bibitem{} Giersz, M., \& Spurzem, R., 2000, MNRAS, 270, 700
\bibitem{}Goodman J., 1983, Ph.D. Thesis, Princeton University
\bibitem{}Goodman J., 1987, ApJ, 313, 576
\bibitem{}Khalisi, E., Doctoral Thesis, Univ. Heidelberg, 2002,
{\tt http://www.ub.uni-heidelberg.de/archiv/3096}
\bibitem{} Kim, E., Einsel, C., Lee, H. M., Spurzem, R. \& Lee M. G.,
   2002, MNRAS, 334, 310
\bibitem{} Kim, S. S., Lee, H. M. \& Goodman, J., 1998, ApJ, 495, 786
 \bibitem{} Lee, H. M., Fahlman, G. \& Richer, H., 1991, ApJ, 366, 455
 \bibitem{} Lee, H. M. \& Goodman, J., 1995, ApJ, 443, 109
 \bibitem{} van Leeuwen, F., Le Poole, R.S., Reijns, R.A., Freeman, K.C., \&
  de Zeeuw, P.T., 2000, A\&A, 360, 471
 \bibitem{} Lupton, R.H., \& Gunn, J.E., 1987, AJ, 93, 1106
 \bibitem{} Portegies Zwart, S.F., Takahashi, K., 1999, CeMDA, 73, 179
   Princeton Univ. Press
 \bibitem{}Spitzer L. Jr. \& Hart M. H., 1971, ApJ, 164, 399
 \bibitem{} Spurzem R., Einsel C., 1998, in D.R. Merritt, M. Valuri, J.A. Sellwood, eds,
   Massive Stellar Clusters, ASP Conf. Ser. No. 182, ASP: San Francisco, p. 105.
 \bibitem{} Takahashi, K., 1995, PASJ, 47, 561
 \bibitem{} Takahashi, K., 1996, PASJ, 48, 691
 \bibitem{} Takahashi, K., 1997, PASJ, 49, 547
 \bibitem{} Takahashi, K. \& Lee, H. M., 2000, MNRAS, 316, 671
 \bibitem{} Takahashi, K., Lee, H. M. \& Inagaki, S., 1997, MNRAS, 292, 331
 \bibitem{} Takahashi, K., \& Portegies Zwart S.F., 1998, ApJL, 503, L49
 \bibitem{} White E.R., Shawl J.S., 1987, ApJ, 317, 246
\end{thebibliography}
\end{document}